\documentclass{JHEP3}
\usepackage{graphics}
\usepackage{graphicx}
\usepackage{amsmath}
\usepackage{cite}
\usepackage{subfig}
\usepackage{tabularx}

\usepackage{amsfonts,epsfig,epstopdf,psfrag,latexsym}

\def\be{\begin{equation}}
\def\ee{\end{equation}}
\def\bea{\begin{eqnarray}}
\def\eea{\end{eqnarray}}
\def\eps{\epsilon}

\def\al{\alpha}

\def\nn{\nonumber}
\def\eps{\epsilon}
\def\Gm{\Gamma}

\def\cO{  {\cal O}  }

\newcommand{\dr}{{\rm d}}

\allowdisplaybreaks[4]

\preprint{TTP13-046}
\title{Evaluating single-scale and/or non-planar diagrams by differential equations}

\author{Johannes M.\ Henn$^{a}$, Alexander V.\ Smirnov$^{b}$, Vladimir A.\ Smirnov$^{c,d}$\\
$^a$ Institute for Advanced Study, Princeton, NJ 08540, USA\\
$^b$ Scientific Research Computing Center, Moscow State University, 119992 Moscow, Russia\\
$^c$ Skobeltsyn Institute of Nuclear Physics of Moscow State University, 119992 Moscow, Russia\\
$^d$ Institut f\"{u}r Theoretische Teilchenphysik, KIT, 76128 Karlsruhe, Germany\\
\email{jmhenn@ias.edu, asmirnov80@gmail.com, smirnov@theory.sinp.msu.ru}}

\abstract{
We apply a recently suggested new strategy to solve differential equations
for Feynman integrals. 
We develop this method further by analyzing asymptotic expansions of the integrals.
We argue that this allows the systematic application of the differential equations 
to single-scale Feynman integrals. Moreover, the information about singular limits
significantly simplifies finding boundary constants for the differential equations.
To illustrate these points we consider two families of three-loop integrals.
The first are form-factor integrals with two external legs on the light cone.
We introduce one more scale 
by taking
one more leg off-shell, $p_2^2\neq 0$.
We analytically solve the differential equations for the master integrals 
in a Laurent expansion in dimensional regularization with $\eps=(4-D)/2$. 
Then we show how to obtain analytic results for the corresponding one-scale integrals in an 
algebraic way. An essential ingredient of our method is to match solutions of
the differential equations in the limit of small $p_2^2$ to our results at $p_2^2\neq 0$ and to identify
various terms in these solutions according to expansion by regions.
The second family consists of four-point non-planar integrals with
all four legs on the light cone.
We evaluate, by differential equations, all the master integrals for the so-called 
$K_4$ graph consisting of four external vertices which are connected with each other by six lines.
We show how the boundary constants can be fixed with the help of the knowledge of the singular limits.
We present results in terms of harmonic polylogarithms for the corresponding seven master integrals with six propagators in
a Laurent expansion in $\eps$ up to weight six. 
}

\keywords{scattering amplitudes, gauge theory, NLO computations,
multiloop Feynman integrals, dimensional regularization, harmonic polylogarithms}

\begin{document}

\section{Introduction}
    
A new strategy of solving differential equations (DE) for Feynman integrals 
\cite{Kotikov:1990kg,Kotikov:1991pm,Remiddi:1997ny,Gehrmann:1999as,Gehrmann:2000zt,Gehrmann:2001ck} was recently suggested \cite{Henn:2013pwa}. 
It is based on choosing a convenient basis of master integrals
that are $\mathbb{Q}$-linear combinations of iterated integrals \cite{Chen1997,Goncharov:1998kja,arXiv:math/0606419} of uniform {\em weight}, 
i.e.  {\em pure} functions of uniform transcendental degree.
The strategy was then successfully applied to the evaluation of
all the three-loop four-point  massless planar diagrams with
all four legs on the light cone  \cite{Henn:2013tua} and to two-loop planar diagrams relevant to 
Bhabha scattering  \cite{Henn:2013woa}. 
In the present paper, we develop this strategy further
and obtain new results at the three-loop level.
  
We pointed out in  \cite{Henn:2013tua} that, as a by-product of the evaluation of 
four-point massless planar diagrams, we also obtained analytic results for planar single-scale three-point
form-factor integrals, although, formally, DE written for one-scale integrals are trivial and
express only the homogeneity of the integrals. Nevertheless, the solution of
a more complicated problem, with one more scale, provided, in a purely algebraic way,
the solution of the one-scale problem, in agreement with the results 
of \cite{Gehrmann:2006wg,Heinrich:2007at,Baikov:2009bg,Heinrich:2009be,Gehrmann:2010ue,Lee:2010ik}.

In this work, the finiteness of planar integrals 
in the $u$-channel as $u=-s-t \to 0$ played a decisive role because these boundary conditions turned out to be
very restrictive. However, in the non-planar case, there are no such simple boundary conditions.
One of the goals of the present paper is to argue that DE, within the strategy of~\cite{Henn:2013pwa}, can
systematically be applied to single-scale Feynman integrals also in this case. 

As we will see, one of the key reasons why this is possible has to do with the fact that the
differential equations contain valuable information about singular limits of Feynman integrals.
To illustrate this, let us take the case of a set $\vec{f}(x,\eps)=\{f_1(x,\eps),\ldots\}$ of master integrals that depend on a 
dimensionless variable $x$ and where $D=4-2 \eps$. 
When applicable, the method of  \cite{Henn:2013pwa} produces a system of differential equations of the Fuchsian type,
\begin{align}\label{DEexact}
\partial_x  \vec{f}(x,\eps) = \eps \, \sum_{i} \frac{A_{i} }{x- x_i} \vec{f}(x,\eps) \,,
\end{align}
with a set of constants $x_i$ and constant matrices $A_{i}$. The perturbative solution in $\eps$ is given by 
iterated integrals built from the alphabet of differential forms $\{ d \log(x-x_{i}) \}$.
Note that at order $k$ in the $\eps$ expansion one has $\mathbb{Q}$-linear combinations of  iterated integrals of uniform weight $k$. 

When one approaches one of the singular points $x_{i}$, which are often of particular
physical interest, $\vec{f}$ has logarithmic singularities.
A typical problem is that the limits $x \to x_{i}$ and $\eps \to 0$ in general do not commute.
Here we point out that the knowledge of eq. (\ref{DEexact}) allows to resolve this order of limits
ambiguity. It is easy to see from  eq. (\ref{DEexact}) that the leading behavior of $\vec{f}$ as $x \to x_i$
is
\begin{align}\label{solsmallx}
\vec{f}(x, \eps) \sim (x-x_{i})^{\eps A_{i}} \vec{g}(\eps) \,,
\end{align}
where $\vec{g}(\eps)=\{g_1(\eps),\ldots\}$ represents the boundary constants. 
Therefore, the singular behavior is governed
by the eigenvalues and eigenvectors of the matrix $A_{i}$.
The crucial point is that eq. (\ref{solsmallx}) allows us to control the non-commutativity of the limits,
since both limits can be generated from the same boundary information $\vec{g}(\eps)$.
In practice, it is often the case that some limit is particularly simple, or can be related to a previously solved problem. In that case, one can determine $\vec{g}(\eps)$ in that limit, and then use it in 
the other limit.\footnote{For three singular points, eq. (\ref{solsmallx}) is a Knizhnik-Zamolodchikov equation \cite{Knizhnik:1984nr}. There, transporting the information from one singular point to another is achieved by the Drinfeld associator.}
This simple observation leads to numerous applications. 
It can be used to determine the asymptotic behavior of Feynman integrals
from fixed-order calculations. 
This applies to physically important singular limits of Feynman integrals and
amplitudes such as threshold expansions, soft limits or Regge limits, to name a few examples.\footnote{In eq. (\ref{solsmallx}), we have only kept the leading term as $x \to x_{i}$. Of course, one can 
also systematically include subleading terms.}
Usually such limits are analyzed using the strategy of expansion by 
regions \cite{Beneke:1997zp,Smirnov:1998vk,Smirnov:1999bza,Smirnov:2002pj},
where the different scalings in eq. (\ref{solsmallx}), corresponding to different eigenvalues
of $A_{i}$, are related to various contributions in asymptotic expansions of Feynman integrals.
The applications we pursue in this paper use the information about limits that is provided by the DE 
to compute single-scale and non-planar integrals.

\begin{figure}[tbp]
  \centering
  \includegraphics[width=0.4\textwidth]{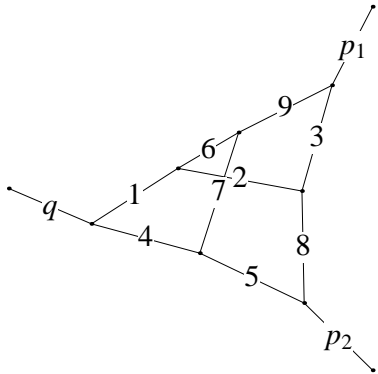}
  \caption{A family of form-factor integrals. Here $-q^{\mu} = p_{1}^{\mu} + p_{2}^{\mu}$, $q^2 = s$ and $p_{1}^2 = 0, p_{2}^2 \neq 0$.}
  \label{figureA92}
\end{figure}

To illustrate our strategy we use the example of a family of
three-loop form-factor master integrals with two external legs on the light cone -- see Fig.~\ref{figureA92}.

We introduce one more scale\footnote{This introduction of one more parameter for single-scale integrals 
in order to use DE was earlier applied in \cite{Schutzmeier:2009}.}
by turning to the corresponding family of integrals with 
one more leg off-shell, $p_2^2\neq 0$,  i.e. depending on two non-zero external momenta squared.
After solving DE for the master integrals we obtain
analytic results in a Laurent expansion in $\eps=(4-D)/2$. Then we show how
to obtain analytic results for the corresponding one-scale integrals in an 
algebraic way. As mentioned above, this is made possible by eq. (\ref{solsmallx}), which allows us 
to match solutions of differential equations in the limit of small $p_2^2$ to our results at $p_2^2\neq 0$.

Another application is to three-loop four-point massless integrals with all four external legs on the light cone. 
We previously computed all planar integrals of this type \cite{Henn:2013tua}, and there are various motivations
for extending this to the non-planar case. It would allow to compute complete three-loop scattering amplitudes
in supersymmetric Yang-Mills and supergravity theories which currently are only known in un-integrated form \cite{Bern:2007hh}.
In particular, this would shed light on the infrared properties of gauge and gravity theories.
Another motivation is to find out whether one can obtain an equation of the form of eq. (\ref{DEexact}), or whether there is
some obstruction due to the non-planar nature of the diagrams.

\begin{figure}[htb]
  \centering
\subfloat[]{\includegraphics[width=0.4\textwidth]{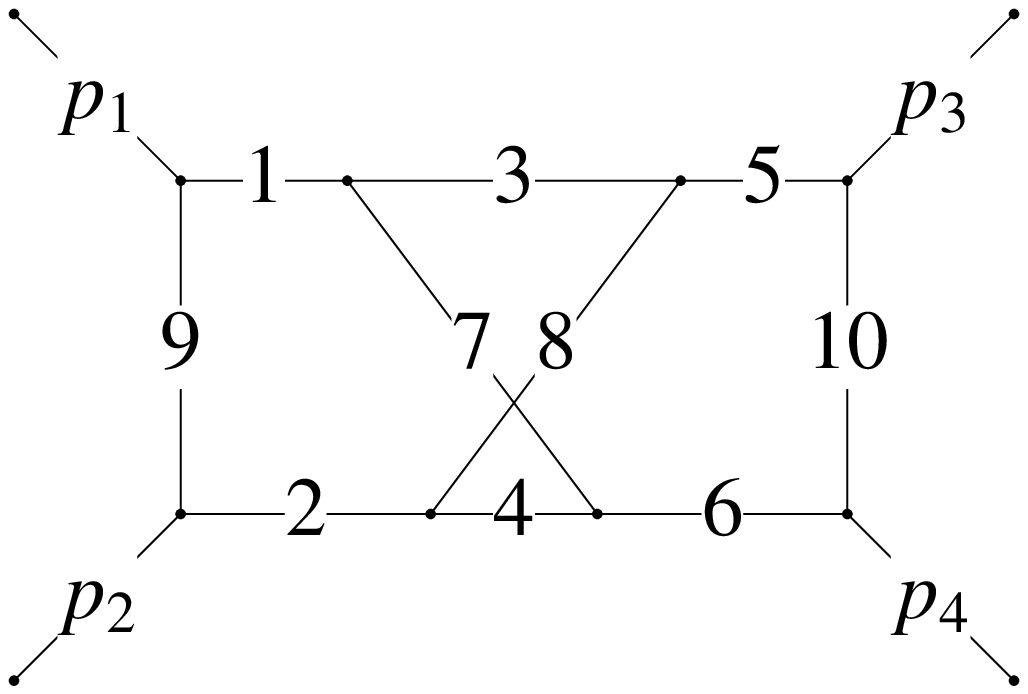}}
\subfloat[]{\includegraphics[width=0.4\textwidth]{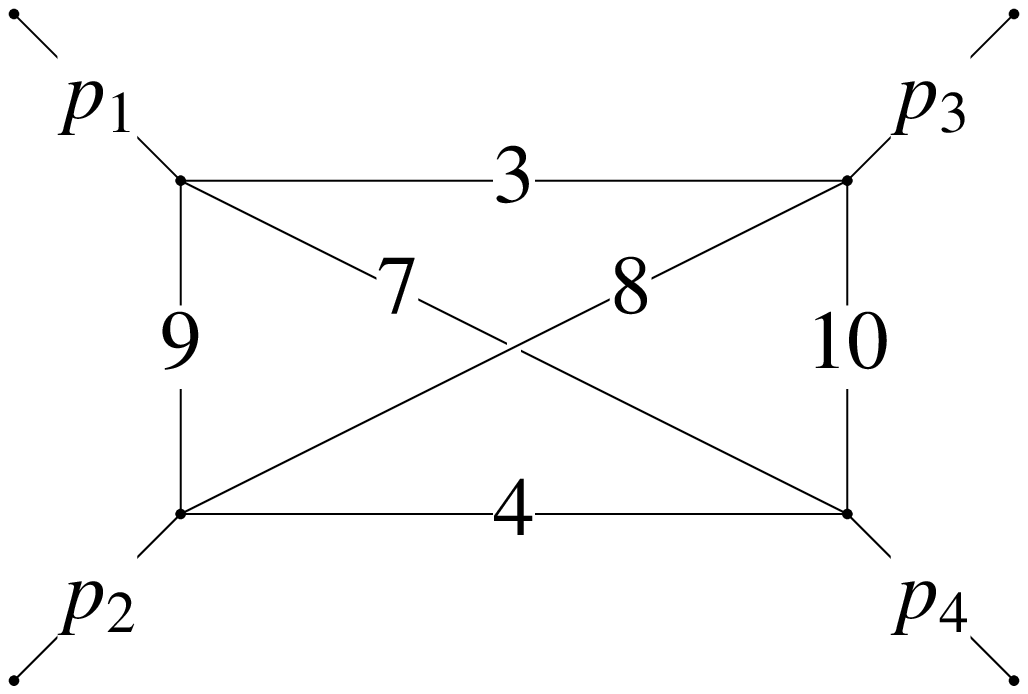}}
    \caption{(a) Diagram C. (b) The $K_4$ graph. All internal lines are massless and $p_{1}^2=p_{2}^2=p_{3}^2=0$. We discuss families of $K_{4}$ integrals for the cases $p_{4}^2=0$ and $p_{4}^2 \neq 0$.}
  \label{K4fig}
\end{figure}

The non-planar case is, however, much more complicated, for various reasons.
Our second goal in the present paper is to evaluate,  within the strategy of~\cite{Henn:2013pwa},
a particularly interesting subfamily of this class  corresponding to the complete four-vertex
graph $K_4$ consisting of four external vertices which are connected with each other by six lines -- see Fig.~\ref{K4fig}(b).
It can be considered as a part of the family~C in \cite{Bern:2007hh}.  
These integrals have fifteen indices: we associate the first ten of them to the edges of
the graph C shown in Fig.~\ref{K4fig}(a) and the last five to numerators. 
Explicitly, we have
\bea
F^C_{a_1,\ldots,a_{15}}(s,t;D) &=&
\frac{1}{(i \pi^{D/2})^3}
\int\int\int \frac{\dr^Dk_1 \, \dr^Dk_2 \, \dr^Dk_3}{(-k_1^2)^{a_1}
[-(p_1 + p_2 + k_1)^2]^{a_2} [-(k_1 + k_3)^2]^{a_3}}
\nonumber \\ && \hspace*{-17mm}
\times \frac{[-(k_1 + k_2)^2]^{-a_{11}}[-(p_1 + k_3)^2]^{-a_{12}} [-(p_1 + k_2)^2]^{-a_{13}} }{
[-(p_1 + p_2 + k_1 + k_2)^2]^{a_4}[-(k_1 + k_2 + k_3)^2]^{a_5}
[-(p_1 + p_2 + k_1 + k_2 + k_3)^2]^{a_6}}
\nonumber \\ && \hspace*{-17mm}
\times \frac{ [-(p_3 + k_1)^2]^{-a_{14}} [-(p_3 + k_3)^2]^{-a_{15}}}{
 (-k_3^2)^{a_7} (-k_2^2)^{a_8} [-(p_1 + k_1)^2]^{a_9}[-(k_1 + k_2 + k_3 - p_3)^2]^{a_{10}} }
\; .
\label{diC}
\eea
Here $s=(p_1+p_2)^2$ and $t=(p_1 + p_3)^2$ denote the Mandelstam invariants and
the causal prescription $-k^2 \longrightarrow -k^2 -i0$ is implied.

For the subfamily associated with the graph $K_4$, we have not only
$a_{11},\ldots,a_{15}\leq 0$ but also $a_{1},a_{2},a_{5},a_{6}\leq 0$.
When we are interested in $K_4$ Feynman integrals only without negative indices, we will
use the notation
\bea
K_{a_1,a_2,\ldots,a_6}= F^C_{0,0,a_1,a_2,0,0,a_3,a_4,a_5,a_6,0,\ldots,0}\;.
\label{K4ints}
\eea
It will be convenient to choose elements of the uniformly transcendental basis
which can have the following non-zero indices: $a_2,a_3,a_4,a_7,a_8,a_9,a_{10}$, with $a_2\leq 0$.
In this case, we will use the notation
\bea
\hat{K}_{a_1,a_2,\ldots,a_6,a'}= F^C_{0,a',a_1,a_2,0,0,a_3,a_4,a_5,a_6,0,\ldots,0}\;,
\label{K4ints1}
\eea
where $a'$ is always non-positive. 
 
The massless graph $K_4$ was recently discussed \cite{Bogner:2013tia} in the context of 
the strategy of evaluating Feynman integrals by iterative integrations over Feynman 
parameters \cite{Brown:2008um}, using multiple polylogarithms.\footnote{See, e.g., \cite{Panzer:2013cha} 
for applications of this strategy, where all three-loop massless
propagator integrals with arbitrary propagator insertions were evaluated up to
$\eps^4$ and some examples at four and more loops were presented.
In particular, a subset of massless four-loop propagator master integrals was evaluated in ref.~\cite{Panzer:2013cha}
in the epsilon expansion up to weight eight, in agreement with the results of ref.~\cite{Baikov:2010hf}
and ref.~\cite{Lee:2011jt} (where results up to weight twelve were presented for all
the master integrals.)}
The graph $K_4$ was discovered not to be linearly 
reducible \cite{Bogner:2013tia}, i.e. it is impossible to find an order of
integration over Feynman parameters such that the dependence on a
current integration parameter would be linear so that every iterative integration
could be performed in terms of multiple polylogarithms.
It was also claimed \cite{Bogner:2013tia} that the presence of $K_4$ as a subgraph is
crucial for the linear irreducibility at higher loops.
The kinematics of the corresponding Feynman integral was considered to be the simplest one
at which the linear irreducibility holds, i.e. all the legs were assumed to be on the light
cone.
 
In the present paper, we show that the $K_4$ Feynman integrals can be
evaluated  in terms of harmonic polylogarithms, in spite of the fact that the graph is linearly 
irreducible. To do this, we apply the strategy of~\cite{Henn:2013pwa} to 
evaluate all the seven master integrals for $K_4$ with six propagators in
a Laurent expansion in $\eps$ up to weight six, which is the typical order for three-loop calculations.

The paper is organized as follows. In section 2, we derive the system of DE satisfied by the class
of non-planar form-factor integrals and discuss the behavior of the integrals in singular limits.
We then use this information to analytically determine all integration constants.
In section 3 and 4, we discuss the family of non-planar four-point $K_{4}$ integrals. 
We present two ways of computing them. In the first method, presented in section 3, we directly derive a system
of differential equations in $x=t/s$ and determine the boundary conditions from symmetry properties and certain asymptotic limits computed via expansion by regions. In the second method (section 4), we introduce one more scale by making one of the external legs off shell. This allows us to fix the integration constants without additional computations, and to solve this three-scale problem analytically.
We conclude in section \ref{sec:discussion}.

Together with the paper, we present ancillary files which contain our results (not only presented in the text) with explanations.

\section{Evaluating single-scale diagrams by differential equations}
\label{sectionasymptoticDE1}
 
\subsection{Analyzing asymptotic behavior with differential equations}

Let us first show how to use differential
equations in order to determine the asymptotic behavior of Feynman integrals.

As an example, we consider a family of massless form-factor integrals with two legs off-shell, see Fig.~\ref{figureA92}.
We have
\begin{align}
G_{a_1 ,\ldots a_{12}} =& \frac{1}{( i \pi^{D/2} )^3}\int\int\int \frac{d^{D}k_1 d^{D}k_2 d^{D}k_3 }
{[-(p_1 + k_{123})^2]^{a_1}[-(p_1 + k_{23})^2]^{a_2}[-(p_1 + k_{3})^2]^{a_3}}  \nonumber \\
& \hspace{-1.5 cm} \times \frac{ [-(p_1+ k_{1})^2]^{-a_{10}}  [-(p_1 + k_{2})^2]^{-a_{11}}  [-( p_2 + k_{3})^2]^{-a_{12}} }{[-(p_2 - k_{123})^2]^{a_4}[-(p_2 - k_{2})^2]^{a_5} [-( k_{1})^2]^{a_6} [-( k_{13})^2]^{a_7}  [-( k_{2})^2]^{a_8}   [-( k_{3})^2]^{a_9} }
\label{A92p2}
\end{align}
where $a_{1}, \ldots a_{9}$ can take any integer values, while $a_{10},a_{11},a_{12}$ correspond to potential numerators and therefore 
can only take non-positive integer values.
Moreover, we use the notation $k_{123} = k_1 +k_2 + k_3$, etc.

 This is a two-scale problem, 
with kinematic invariants $(p_1 + p_2)^2 =s$ and $p_2^2$, while $p_1^2=0$. We denote the dimensionless ratio by $x= p_2^2/s$. 

Let us write down the $\alpha$ (or Feynman) 
representation\footnote{Eq. (\ref{alpha-d-mod}) is for $a_{10}=a_{11}=a_{12}=0$. 
Let us note that integrals with numerators (negative indices) can be considered
by the same formula (\ref{alpha-d-mod}) where auxiliary $\al$-parameters are introduced for
the negative indices. A differentiation of order $-a_i$ for such indices is implied and then
they are set to zero, so that a resulting $\alpha$-parametric integral has the same structure as
with nine positive indices but the powers of the two basic functions become shifted and
an extra polynomial in the integrand appears.
This extra polynomial comes from the differentiation and therefore is only in the numerator and does not alter 
the analytic structure.}
\bea
\frac{
\Gm(a-h D/2)}{\prod_l\Gm(a_l)}
\int_0^\infty d\al_1 \ldots\int_0^\infty d\al_L \,
\delta\left( \sum \al_l-1\right)
\frac{
{\cal U}^{a-(h+1) D/2} \prod_l \al_l^{a_l-1} }
{ ({\cal W}-i 0)^{a-h D/2}} \;,
\label{alpha-d-mod}
\eea
where $a=\sum_{i=1}^{L} a_{i}$, $h=3$ is the number of loops and ${\cal U}$ and ${\cal W}$ are basic polynomials which are given by well-known
graph-theoretical formulae, see e.g. \cite{Smirnov:2012gma}.
The main point is that
\begin{align}
\cal{W} &= (-s) \,W_s + (-p_{2}^2) \, W_{p_{2}^2} \,.
\end{align}
where $W_s$ and $W_{p_{2}^2}$ are positive polynomials in the $\alpha_{i}$.
From this and eq. (\ref{alpha-d-mod})
it follows that all integrals are real when $s<0, p_{2}^2<0$, i.e. for $x>0$.
The same fact allows us to absorb the $i0$ from the Feynman prescription of the propagators in the kinematical variables, $-s \to -s-i0$ and $-p_{2}^2 \to -p_{2}^2 -i0$. Setting $s=-1$ without loss of generality, this means that $x$ acquires a small negative imaginary part, $x \to x - i0$. We will leave this implicit in the formulas presented below.

Using IBP relations with the help of the {\tt c++} version of {\tt FIRE} \cite{Smirnov:2008iw,Smirnov:2013dia}, 
we find that this family can be spanned by a basis of $39$ master integrals $\vec{f} = \{f_1,f_2,\ldots,f_{39}\}$. 
For example, one of the most difficult nine-propagator integrals we take as basis elements is
\begin{align}
f_{38} = \eps^6 (p_2^2 - s)^2 (-s)^{3 \eps} G_{1,1,1,1,1,1,1,1,1,0,-1,0} \,.
\end{align}
We have normalized all integrals such that they are dimensionless functions, and so that their $\eps$ expansion starts at $\eps^0$. In particular, they cannot have branch cuts starting at $x=1$, and this
information will be useful when determining boundary constants for the integrals.

We find the following system of differential equations
\begin{align}\label{DE}
\partial_x \, \vec{f}(x,\eps) = \eps \left[ \frac{A}{x} + \frac{B}{x-1} \right] \, \vec{f}(x,\eps) \,,
\end{align} 
where $A$ and $B$ are constant $39 \times 39$ matrices.
The singular points $x= 0,1,\infty$ of eq. (\ref{DE}) correspond to the on-shell limit $p_2^2=0$, the two-point function limit $p_1 = 0$, and the on-shell limit $s=0$.
It turns out that $x \to 1$ is an excellent limit for determining boundary conditions, because many integrals either vanish or are known simple functions
at that point. For a few integrals we have also used the limit $x\to \infty$. These limits, together with simple analytic expressions for propagator-type integrals
that are known in terms of gamma functions, completely determine the boundary constants, and therefore allow us to obtain the full solution.

The solution at any order $\eps^k$ is given by a linear combination (with rational coefficients) of harmonic polylogarithms 
$H_{a_1,a_2,\ldots,a_n}(x)$ \cite{Remiddi:1999ew} of weight $k$.
The latter are iterated integrals built from the alphabet of differential forms $d\log x, d\log(1-x), d\log(1+x)$.
More precisely,
\begin{equation}
H_{a_1,a_2,\ldots,a_n}(x) = \int_0^x  f_{a_1}(t) H_{a_2,\ldots,a_n}(t)\,\dr t\;,
\label{HPL-def}
\end{equation}
where
\bea
f_{\pm 1}(x)&=&\frac{1}{1 \mp x}\;,\;\;f_0(x)=\frac{1}{x}\;,
\label{HPL-f-def}
\\
H_{\pm 1}(x)&=& \mp \log(1\mp x),\;\; H_0 (x)= \log x \;,
\label{HPL-H01-def}
\eea
and at least one of the indices $a_i$ is non-zero.
For all $a_i=0$, one has
\begin{equation}
H_{\underbrace{0,0,\ldots,0}_{n}}(x) = \frac{1}{n!}\log^n x\;.
\end{equation}

{}From eq. (\ref{DE}) we see that
in fact only the first two letters of this alphabet are required, or, in other words, only the indices $0$ and $+1$. 
We explicitly expanded the solution to weight eight. In the remainder of this section, for the sake of readability, we will often truncate formulas
at some lower order in $\eps$.

To evaluate massless form-factor integrals, let us consider the limit $x\to 0$. 
This limit does not in general commute with the $\eps$ expansion, and therefore naively 
one cannot use the result for fixed order in $\eps$ to compute the massless form-factor integrals.
However, the missing information can be obtained by the differential equation (\ref{DE}). 
It is easy to solve it for small $x$ and for finite $\eps$, by neglecting the $B/(x-1)$ term. 
The solution in that regime takes the form
\begin{align}
\vec{f}(x, \eps) \sim x^{\eps A} \vec{g}(\eps)\,,
\end{align}
where $g(\eps)$ are boundary constants, to be determined.
Note that $x^{\eps A}$ is a matrix exponential, which can easily be computed for a given constant matrix $A$. 
In a typical situation, where the matrix $A$ is non-diagonalizable,
we find
 $x^{-\alpha_{j} \eps} \log^k(x)$, where $\alpha_{j}$ are eigenvalues of $A$. 
So, the solution in that regime looks like 
\begin{align}\label{DEasy}
f_{i} \sim \sum_{j,k} c_{ijk}(\eps) \, x^{-\alpha_{j} \eps} \log^k(x) \,,
\end{align} 
with the $\alpha_{j}$ are eigenvalues of $A$, and the $c_{ijk}(\eps)$ are determined by $g(\eps)$. 
(One could make this relationship more concrete by referring to the eigenvectors and in general
power vectors of the matrix $A$.)
Expanding this formula for small $\eps$, we can determine the matching coefficients $c_{ijk}$ 
by comparing to our results in a Laurent expansion in $\eps$. 
Then, we can return to formula (\ref{DEasy}), keep only the terms with $\alpha_{j}=0$
and, therefore arrive at the form-factor integrals with the external momentum $p_2$ on-shell, i.e. $x=0$.

Indeed, the integrals considered on-shell, or at a threshold are, by definition, obtained from
integrals at general values of a given external momentum by setting it on-shell or a threshold 
under the integral sign either in integrals over loop momenta or in parametric integrals.
On the other hand when we consider the limit $x\to 0$ we can apply the strategy of expansion by 
regions~\cite{Beneke:1997zp,Smirnov:1998vk,Smirnov:1999bza,Smirnov:2002pj} (see also Chapter~9 of \cite{Smirnov:2012gma}).
According to this strategy, the expansion in a given limit is given by a finite number of series corresponding
to so-called regions which are scalings of certain components of the loop momenta in terms of a given
parameter of expansion, e.g. $x$. One of the regions corresponds to all the components of the loop momenta (or all the
parameters in alpha representation) to be of order $x^0$. It is usually called hard. Its contribution is given by Taylor expanding
the integrands in $x$. The leading term in this contribution is obtained just by setting $x=0$ under integral sign.
Obviously, this contribution corresponds to $\alpha_{j}=0$ in eq. (\ref{DEasy}).
Other regions typical for Sudakov and Regge limits are called collinear (with $\alpha_{j}=1$ per loop) and ultrasoft (with $\alpha_{j}=2$  per loop).

\subsection{Example}

As an example of this, let us discuss the solution for one of the most non-trivial integrals, $f_{38}$. Solving the system of differential equations with the appropriate boundary conditions as discussed above, we find
\begin{align}\label{res38x}
f_{38}(x) = \eps^3\, \left[ \frac{1}{9} \pi^2 H_{0}(x) + \frac{4}{3} H_{0,0,0}(x) + \frac{2}{3} H_{0,1,0}(x) + \frac{4}{3} \zeta_{3} \right] + \cO(\eps^4) \,.
\end{align}

Equation (\ref{res38x}) is the result for integral $f_{38}$ in the small $\eps$ limit. When taking in addition $x\to0$, divergences appear from the logarithms in that formula. As explained above, we can understand these divergences from the general solution of eq. (\ref{DE}), namely eq. (\ref{DEasy}). In the present case, we find\footnote{We remark that all integrals in 
our basis are UV finite, but in general have IR divergences, so that we can consider $\eps<0$ and finite.}
\begin{align}\label{asy38}
\lim_{x \to 0} f_{38}(x) \sim c_0(\eps) + c_1(\eps) x^{-\eps} + c_2(\eps) x^{-2 \eps} + c_3(\eps) x^{-3 \eps}\,.
\end{align}
For $\eps \to0$, the $x^{-\alpha \eps}$ terms lead to the logarithms we observed above.
We can also see that there is an order of limits issue when considering $\eps \to 0$ and $x\to 0$. 

On the other hand,  we arrive at this general structure from the analysis of the system of differential equations.
Moreover, we determine the expansion coefficients $a_{i}(\eps)$ by comparing against the solution for small $\eps$. 
Indeed, let us compare the small $x$ limit of eq. (\ref{res38x}) to the small $\eps$ limit of eq. (\ref{asy38}). This leads to a constraint on the $c_{i}(\eps)$. 
Of course, we have an analog of eq. (\ref{asy38}) for all integrals, which implies that we can completely determine the coefficients $a_{i}(\eps)$  by this procedure. 
For notational brevity, we present the results only up to order $\eps^4$,
\begin{align}
c_{0} =& \frac{2}{9} - \frac{17}{54}\pi^2 \eps^2 - \frac{31}{3} \zeta_{3} \eps^3 - \frac{119}{432} \pi^4 \eps^4 + \cO(\eps^5)\,, \\
c_{1} =&-\frac{2}{3} + \frac{8}{9}\pi^2 \eps^2 + \frac{100}{3} \zeta_{3} \eps^3 + \frac{2827}{2160} \pi^4 \eps^4 + \cO(\eps^5)\,,\\
c_{2} =&\frac{2}{3} - \frac{13}{18}\pi^2 \eps^2 - \frac{103}{3} \zeta_{3} \eps^3 - \frac{3149}{2160} \pi^4 \eps^4 + \cO(\eps^5)\,,\\
c_{3} =&-\frac{2}{9} + \frac{4}{27}\pi^2 \eps^2 + \frac{38}{3} \zeta_{3} \eps^3 + \frac{1133}{2160} \pi^4 \eps^4 + \cO(\eps^5)\,.
\end{align}
Having obtained these `matching coefficients', we can return to eq. (\ref{asy38}) and consider the opposite order of limits, 
i.e. $x\to0$ for finite (negative) $\eps$. 
In this case, we ignore the terms $x^{-\alpha \eps}$ with positive $\alpha$  and we are left with
\begin{align}
f_{38}(x=0) = c_{0}(\eps) \,.
\end{align}
This is nothing but the value for the massless form-factor integral.
Comparing to eq. (22) of ref. \cite{Lee:2010ik}, we find perfect agreement to order $\eps^8$. 
We have calculated, in a similar way, all the master integrals for the family of Feynman integrals (\ref{A92p2})
at $p_1^2=p_2^2=0$ and found agreement with the results \cite{Lee:2010ik} up to weight eight and earlier results up
to weight six~\cite{Heinrich:2009be,Baikov:2009bg}.
We stress that the calculation performed here was done entirely from first principles, using only algebraic steps. 

Let us comment on the general structure of the asymptotic expansion.
The $x^{-\alpha \eps}$ terms correspond to contributions of certain regions \cite{Beneke:1997zp}.
Although we did not carry out a detailed analysis, one would expect that 
the term $x^{- \eps}$ corresponds to regions where one of the loop momenta is collinear and two loop momenta
are hard, the term $x^{-2\eps}$ corresponds to regions where two loop momenta are collinear and one loop momentum is is
hard, and the term $x^{-3\eps}$ corresponds to regions where all the loop momenta are collinear.
As discussed above, the term $x^{0\cdot\eps}$ corresponds to the region of the three hard momenta.
Although intuitive, in general it is hard to find these regions in momentum space systematically.
In contrast, revealing regions in the space of Feynman/alpha parameters \cite{Smirnov:1999bza,Smirnov:2012gma} can be made automatic.
To do this, one can apply an open code {\tt asy.m} \cite{Pak:2010pt,Jantzen:2012mw}.
In this particular example, this code reports about seven contributions corresponding to certain regions.
The power dependence on $x$ in these contributions exactly  corresponds to the exponents present  in (\ref{asy38}).
More specifically, there is one region with $x^{0\cdot\eps}$ (this always takes place), one region with $x^{-\eps}$,
two regions with $x^{-2\eps}$, and  three regions with $x^{-3\eps}$.

\section{Evaluating $K_{4}$ integrals} 
\label{sectionK4onshell}

Here we evaluate the $K_{4}$ integrals defined in the introduction, see eq. (\ref{diC}) and Fig.~\ref{K4fig}(b).
As mentioned in the introduction, these integrals are a non-planar version of the three-loop integrals solved for in \cite{Henn:2013tua}. In that reference, 
we used a simple boundary condition of the absence of singularities in the $u$-channel in order to determine the boundary constants. For non-planar integrals, 
we do not have a similar condition. However, the setup discussed in section \ref{sectionasymptoticDE1} gives us control even in the case where the limits are singular, and this will help us in fixing the integration constants.

There is another complication related to the non-planar nature  of the integrals
that can be seen by looking at the $\alpha$ representation.
The polynomials $\cal U$ and $\cal W$ in eq. (\ref{alpha-d-mod}) are given by
\begin{align}
{\cal U} =
& \al_1 \al_2 \al_3 + \al_1 \al_2 \al_4 + \al_1 \al_3 \al_4 + \al_2 \al_3 \al_4 + 
 \al_1 \al_2 \al_5 + \al_2 \al_3 \al_5 + \al_1 \al_4 \al_5 + \al_3 \al_4 \al_5 \nn
\\ 
& + 
 \al_1 \al_2 \al_6 + \al_1 \al_3 \al_6 + \al_2 \al_4 \al_6 + \al_3 \al_4 \al_6 + 
 \al_1 \al_5 \al_6 + \al_2 \al_5 \al_6 + \al_3 \al_5 \al_6 + \al_4 \al_5 \al_6 \;, \label{defUK4} \\
{\cal W}= &(-s) \al_1 \al_2 \left(\al_3 \al_4-\al_5 \al_6 \right) + (-t) \al_5 \al_6 \left(\al_3 \al_4-\al_1 \al_2 \right) \label{defWK4}\,.
\end{align}
We see from eq. (\ref{defWK4}) that ${\cal W}$ does not have a definite sign (for some region of $s,t$), and as a consequence we cannot treat the $i0$ prescription in eq. (\ref{alpha-d-mod}) simply as a complex deformation of $s$ and $t$.
This can also be seen simply by looking at bubble integrals in the 
$s$, $t$ and $u$-channel, which give rise to logarithms
\begin{align}
\log(-s-i0) \,,\qquad \log(-t-i0) \,,\qquad \log(-u-i0) = \log(-s-t+i0) - i \pi \,,
\end{align}
where we used $u=-s-t$. We therefore need to be careful about those different $i0$ prescriptions.
As in section \ref{sectionasymptoticDE1}, we define dimensionless functions of one variable,
$x=t/s$. In the calculation below we will assume $x>0$, unless otherwise stated.

Solving IBP relations \cite{Chetyrkin:1981qh} with the help of the {\tt c++} version of {\tt FIRE} \cite{Smirnov:2008iw,Smirnov:2013dia}, 
for the family of integrals (\ref{diC}),
we find that, at   $a_{1},a_{2},a_{5},a_{6},a_{11},\ldots,a_{15}\leq 0$,
there are three trivial master integrals with four propagators and seven master 
integrals\footnote{We checked this number with the help of a recently published code \cite{Lee:2013hzt}
which is based on the analysis of \cite{Smirnov:2010hn}.}
with six propagators.
In different situations, it is reasonable to choose different bases of the master integrals.

One can try to evaluate the master integrals by 
the method of Mellin-Barnes (MB) representation \cite{Smirnov:1999gc,Tausk:1999vh,Smirnov:2012gma}. 
Since it is usually complicated to derive MB representation with general numerators,
so it is better to choose master integrals with $a_{1}=a_{2}=a_{5}=a_{6}=a_{11}=\ldots=a_{15}= 0$
and six positive indices. For example, one can choose
$K_{1,1,1,1,1,1}$,$K_{2,1,1,1,1,1}$,$K_{1,1,2,1,1,1}$,$K_{1,1,1,1,2,1}$,$K_{2,2,1,1,1,1}$,$K_{1,1,2,2,1,1}$, and $K_{1,1,1,1,2,2}$ as an alternative basis.

An eight-fold MB representation for $K_4$-integrals without numerator is presented in Appendix~A.
However, the  straightforward procedure of evaluating these integrals by the MB representation
works well only up to weight three in the epsilon-expansion, while we have an implicit
obligation to obtain results up to weight six. Therefore, we will now use differential equations, and
the results obtained with the MB representation will be used for checks.

We will use a similar notation as in section 2, hoping that this will not lead to confusion.
After deriving differential equations for this class of integrals in the kinematical variables $s$ and $t$, we use the freedom to choose a convenient integral basis. 
The choice we made is 
\begin{align}\label{K4rescaling}
\vec{f} = e^{3 \eps \gamma_{E}}  (-s)^{-3\eps}  \vec{g}\,, 
\end{align}
with
\begin{align}\label{K4basischoice}
g_1=& \eps^3 t K_{0, 0, 1, 2, 2, 2}, \qquad
g_2= \eps^3 (s + t) K_{1, 2, 0, 0, 2, 2} , \\
g_3=& \eps^3 s K_{1, 2, 2, 2, 0, 0},\qquad
g_4= 2 \eps^4 (s+t) \hat{K}_{1, 2,  1, 1, 2, 1,-1} + 
  2 \eps^5 s K_{2, 1,   1, 1, 1, 1} ,\\
   g_5=&4 \eps^5 t K_{2, 1,  1, 1, 1, 1} ,\qquad
   g_6=4 \eps^5 (s+t) K_{1, 1,  2, 1, 1, 1} ,\\
   g_7=& 4 \eps^5 s  K_{1, 1, 1, 1, 2, 1} ,\qquad
   g_8= - 2 \eps^4 s (s+t) K_{2, 2, 1, 1, 1, 1},\\
   g_9=& -2 \eps^4 s t K_{1, 1, 2, 2, 1, 1} ,\qquad
   g_{10}= -2 \eps^4 (s+t) t K_{1, 1, 1, 1, 2, 2} ,  
\end{align}
where we use (\ref{K4ints}) and (\ref{K4ints1}).
As we will see presently, these functions will be pure functions of uniform weight.

For the basis choice of eq. (\ref{K4basischoice}) we find the system of DE,
\begin{align}\label{DEK4}
\partial_x \vec{f}(x,\eps) = \eps \left[ \frac{A}{x} + \frac{B}{1+x} \right] \vec{f}(x,\eps) \,.
\end{align}
This is the same form previously found for planar three-loop integrals in \cite{Henn:2013tua}.
Here $A$ and $B$ are the following constant $10 \times 10$ matrices\,,
\begin{align}
A = \left(
\begin{array}{cccccccccc}
 -3 & 0 & 0 & 0 & 0 & 0 & 0 & 0 & 0 & 0 \\
 0 & 0 & 0 & 0 & 0 & 0 & 0 & 0 & 0 & 0 \\
 0 & 0 & 0 & 0 & 0 & 0 & 0 & 0 & 0 & 0 \\
 -\frac{2}{3} & \frac{2}{3} & -\frac{1}{6} & 1 & \frac{1}{3} & -\frac{1}{3} & -\frac{7}{6} & \frac{1}{12} & -\frac{1}{12} & \frac{1}{3} \\
 0 & 0 & 0 & 0 & 1 & 0 & 0 & 0 & 0 & 0 \\
 -1 & 1 & 1 & 4 & 5 & -3 & -3 & -\frac{1}{2} & 0 & \frac{1}{2} \\
 \frac{1}{3} & \frac{5}{3} & \frac{1}{3} & 4 & \frac{7}{3} & -\frac{7}{3} & -\frac{11}{3} & -\frac{1}{6} & \frac{1}{6} & \frac{1}{3} \\
 -\frac{4}{3} & \frac{10}{3} & -\frac{10}{3} & 0 & \frac{20}{3} & \frac{10}{3} & -\frac{10}{3} & \frac{5}{3} & -\frac{2}{3} & \frac{2}{3} \\
 -\frac{14}{3} & \frac{8}{3} & \frac{4}{3} & 8 & \frac{22}{3} & -\frac{16}{3} & -\frac{20}{3} & -\frac{2}{3} & -\frac{7}{3} & \frac{4}{3} \\
 \frac{10}{3} & \frac{8}{3} & \frac{4}{3} & 8 & \frac{22}{3} & -\frac{16}{3} & -\frac{20}{3} & -\frac{2}{3} & \frac{2}{3} & -\frac{5}{3}
\end{array}
\right)\,,
\end{align}
and
\begin{align}
B = \left(
\begin{array}{cccccccccc}
 0 & 0 & 0 & 0 & 0 & 0 & 0 & 0 & 0 & 0 \\
 0 & -3 & 0 & 0 & 0 & 0 & 0 & 0 & 0 & 0 \\
 0 & 0 & 0 & 0 & 0 & 0 & 0 & 0 & 0 & 0 \\
 \frac{1}{6} & -\frac{1}{6} & \frac{1}{6} & -5 & -\frac{7}{3} & \frac{5}{6} & \frac{1}{6} & \frac{1}{6} & \frac{1}{12} & -\frac{1}{12} \\
 -\frac{1}{3} & \frac{1}{3} & -\frac{1}{3} & 4 & \frac{5}{3} & \frac{1}{3} & -\frac{1}{3} & -\frac{1}{3} & -\frac{1}{6} & \frac{1}{6} \\
 0 & 0 & 0 & 0 & 0 & 1 & 0 & 0 & 0 & 0 \\
 -\frac{1}{3} & -\frac{5}{3} & -\frac{1}{3} & -4 & -\frac{7}{3} & \frac{7}{3} & -\frac{1}{3} & \frac{1}{6} & -\frac{1}{6} & -\frac{1}{3} \\
 0 & -2 & 0 & 8 & 4 & -2 & 0 & -3 & 0 & 0 \\
 \frac{10}{3} & -\frac{4}{3} & \frac{10}{3} & 0 & \frac{10}{3} & \frac{20}{3} & \frac{10}{3} & -\frac{2}{3} & \frac{5}{3} & -\frac{2}{3} \\
 0 & -6 & 0 & -8 & -4 & 2 & 0 & 0 & 0 & -3
\end{array}
\right) \,.
\end{align}
It is clear from the discussion in section \ref{sectionasymptoticDE1} that we can solve for
the functions $\vec{f}(x,\eps)$ in an expansion in $\eps$, where the expansion coefficients are given by harmonic polylogarithms. In the next subsection, we discuss how we used information about the asymptotic behavior at singular points in order to determine the boundary constants.\footnote{In section \ref{sectionK4offshell}, we will see an alternative approach that does not require an additional calculation in order to determine boundary conditions, in the same spirit as section \ref{sectionasymptoticDE1}.}

\subsection{Asymptotic behavior and boundary conditions}
\label{sectionK4limits}

The first three functions $g_1, g_2 , g_3$, are simple functions that can be given analytically
in terms of $\Gamma$ functions. Therefore we only need to specify boundary values for the
remaining seven functions.  

As discussed above, the eigenvalues of $\eps A$, $\eps B$, and $\eps (A+B)$ 
characterize the three singular limits of $\vec{f}$.
We have the eigenvalues $\{ -3 \eps  ,0, \eps \}$ for $t \to 0$ and $\{ -4 \eps , -3 \eps , 0 \} $
for $s \to 0$.  (Notice the rescaling by $(-s)^{-3 \eps}$ in eq. (\ref{K4rescaling}).)
Finally, for $u \to 0$, we have $\{ -3 \eps ,0, \eps \}$.
We see that some of the eigenvalues are positive, e.g. $\eps A$ has eigenvalues $+\eps$.
This is slightly surprising, for the following reason. All integrals in our basis $\vec{f}$ are UV finite, and IR divergent. 
Therefore, they can be defined for $\eps<0$, and in particular they should stay finite if we take a limit 
such as  $x \to 0$ (with $\eps$ finite). However, a term like $x^\eps$ corresponding to the eigenvalue $+ \eps$ 
would diverge when $x\to 0$. Therefore we expect that the coefficients of such terms must vanish. 
This requirement fixes some of the integration constants.

In order to have further analytic boundary conditions, we computed asymptotic expansions
using the computer code {\tt asy.m} \cite{Pak:2010pt,Jantzen:2012mw} which is now included in {\tt FIESTA3} \cite{FIESTA3}.
This code uses the information about the propagators of a given Feynman integral as an input
and produces the corresponding set of regions relevant to a given limit, in the language of Feynman
parameters. The search of regions reduces to finding faces of maximal dimension
of the Newton polytope associated with the two basic polynomials in the alpha representation.
This code provides various contributions to a given asymptotic expansion as parametrical integrals
and performs as many explicit integrations in these integrals as possible.
In the case of the limit $t\to 0$, these are the contributions (called hard and collinear) characterized by exponents 
$x^0$ and  $x^{ -3 \eps}$   of the expansion parameter, in agreement with the discussion above.
Starting from these parametric integrals we derived a one-fold MB representation
for the collinear-type contributions (with the exponent $x^{ -3 \eps}$).
The evaluation of the corresponding MB integrals in the $\eps$-expansion
is straightforward. It reduces to summing up one-fold series, which can be done with the help of
public computer codes \cite{Vermaseren:1998uu,Moch:2005uc}.
The results obtained can be expressed in terms of multiple zeta values. 
 
Let us illustrate this procedure using the master integral $K_{2,2,1,1,1,1}$.
The code {\tt asy.m} reveals one hard contribution and two collinear contributions. To deal with
the two collinear contributions individually, one has to introduce an auxiliary analytic regularization.
This can be done by shifting the first index $2$ by an analytic parameter $\lambda$, i.e. we
will consider $a_1=2+\lambda$. Then the code {\tt asy.m} produces the following expression
for the sum of the two collinear contributions:
\bea
x^{-3 \eps}\frac{ \Gamma (-4 \eps) \Gamma (3\eps)\Gamma (-\lambda) }{\Gamma (-4 \eps-\lambda)} \int_0^1\ldots\int_0^1 d \al_1\ldots d \al_4
\delta\left(\sum_i \al_i-1\right)
\al_1^{-3 \eps} \al_2^{-3 \eps} \al_3^{-3\eps} \al_4^{-3 \eps} 
&& \nn \\ &&  \hspace*{-135mm} 
\times
(\al_1 \al_2 \al_3+\al_1 \al_4 \al_3+\al_2 \al_4 \al_3+\al_1
   \al_2 \al_4)^{4 \eps} (\al_2+\al_3)^{\lambda} (\al_1+\al_4)^{\lambda} 
   (\al_1 \al_2-\al_3\al_4-i0)^{-\lambda-2}
\nn \\ &&  \hspace*{-135mm} 
+
x^{-3\eps-\lambda} 
\frac{\Gamma (-4 \eps-2 \lambda) \Gamma (3\eps+\lambda)\Gamma(\lambda)}{\Gamma (\lambda+2) \Gamma (-4\eps-\lambda)}
\int_0^1\ldots\int_0^1 d \al_1\ldots d \al_4
\delta\left(\sum_i \al_i-1\right)
\nn \\ &&  \hspace*{-135mm}   
\times
   \al_1^{-3 \eps-\lambda} \al_2^{-3 \eps-\lambda} \al_3^{-3\eps-\lambda} \al_4^{-3 \eps-\lambda} 
   (\al_1 \al_2 \al_3+\al_1 \al_4 \al_3+\al_2 \al_4 \al_3+\al_1 \al_2 \al_4)^{4 \eps+2 \lambda}   
\nn \\ &&  \hspace*{-135mm}   
\times   
(\al_1+\al_3)^{-\lambda} (\al_2+\al_4)^{-\lambda}    
(\al_1 \al_2-\al_3 \al_4-i0)^{-2} \;.
\eea
As is well known, one can choose any subset of the parameters in the argument of the delta functions
involved.

The second of the two integrals can be evaluated as follows. We choose the argument of the
delta function as $\al_4-1$ so that we set $\al_4=1$ and obtain an integral from $0$ to $\infty$ over
the remaining three parameters. 
We turn to the new variables by $\al_1= \eta \xi, \al_3= \eta (1-\xi)$ and 
integrate explicitly over $\eta$. Then we separate the two terms in $(-(1 - \xi) + \xi \al_2-i0)$
by introducing a MB integration and take explicitly integrations over $\al_2$ and $\xi$
in terms of gamma functions. One can proceed similarly with the first integral.
Then we can take the limit $\lambda\to 0$ in the sum of the two integrals to obtain
the following expression for the sum of the two collinear contributions in terms of a one-fold
MB integral:
\bea
 - x^{-3 \eps} \frac{\Gamma (3 \eps)}{\Gamma (-4 \eps)}
 \int_{-\infty}^{+\infty} d z\;  e^{i \pi  z} \Gamma (-z) \Gamma(z+2) \Gamma (-\eps-z-1)^2 \Gamma (-\eps+z+1)^2
&& \nn \\ &&  \hspace*{-120mm} 
\times 
 \left(4 \psi(-2 \eps)-\psi(3 \eps)+\log(x)-2 \psi(-\eps-z-1)+\psi(z+2)+i \pi +2 \gamma_{\rm E} \right) \;.
\eea
One can evaluate this integral in a Laurent expansion in $\eps$ by the standard 
procedures \cite{Czakon:2005rk,Smirnov:2009up,heptools1,heptools2},
first, resolving the singularities of the integral in $\eps$ and then converting the integrals
obtained into a series and summing it up \cite{Vermaseren:1998uu,Moch:2005uc}.

We obtain the following result for the sum of the leading order collinear contributions to $K_{2,2,1,1,1,1}$
in the limit $x\to 0$:
\bea
 x^{-3 \eps} \left[-\frac{421}{5} \zeta_5 \log
   (x)+\frac{29}{12} \pi ^2 \zeta_3 \log (x)-\frac{421 i
   \pi  \zeta_5}{10}+\frac{5597 \zeta (3)^2}{36} \right.
&& \nn \\ &&  \hspace*{-90mm}    
\left.
   +\frac{29}{24} i \pi ^3
   \zeta_3+\frac{31601 \pi ^6}{2177280} + O(x)\right]\;.
\eea
For the hard contribution (i.e. terms with the exponent $x^0$), it was possible to derive a four-fold
MB representation. Moreover, we could simplify the evaluation taking into account the fact
that the corresponding contributions are given by Feynman integrals depending on two, rather than
three external momenta because the kinematics $t=0$ implies $p_3=-p_1$.
Therefore, we could apply an IBP reduction to such integrals.
The number of the corresponding master integrals drastically reduces: in the sector with
six positive indices, it is equal to two instead of seven.
We used the four-fold MB representation to evaluate these master integrals in lower orders
of the $\epsilon$ expansion. However, as we will discuss in the next subsection,
we obtained sufficient boundary information in other ways, 
so that we used such results only for checks.

\subsection{Crossing symmetry}
\label{section:crossing}

The $K_{4}$ integrals have a huge amount of symmetry under exchange of external momenta.
For example, studying the exchanges $p_{1} \leftrightarrow p_{2}$ and $p_{1} \leftrightarrow p_{3}$, we find the following relations, 
\begin{align}
f_{6}(x) = -f_{5}(-1-x)\,, \qquad f_{9}(x) = -f_{8}(-1-x)\,,
\end{align}
and
\begin{align}
f_{7}(x) = x^{-3 \eps} f_{5}(1/x) \,, \qquad f_{10}(x) = x^{-3 \eps} f_{8}(1/x) \,,
\end{align}
and similarly for $f_{1}, f_{2}, f_{3}$.
This can be seen by inspecting the $\alpha$ representation, and in particular the
polynomial $\cal W$ of eq. (\ref{defWK4}) for those integrals. 
In this way one sees that the relations that interchange $x$ and $-1-x$ are valid for arbitrary
real $x$, while the relations that interchange $x \leftrightarrow 1/x$ above are only valid for $x>0$. For $x<0$, the $i0$ prescription leads to the following modification,
\begin{align}
f_{7}(x) = \left[ (x+i0)^{-3 \eps} f_{5}(1/x) \right]^{*} \,, \qquad  f_{10}(x) = \left[(x+i0)^{-3 \eps} f_{8}(1/x)  \right]^{*} \,, \qquad   x<0\,,
\end{align}
where ${}^{*}$ stands for complex conjugation and $\eps$ is supposed to be real.


We used these symmetries in order to fix the boundary constants remaining from the discussion in subsection \ref{sectionK4limits}. 
The remaining relations served as a check of our calculation.

\subsection{Result for $K_{1,1,1,1,1,1}$}

Using the results for the boundary conditions of subsection (\ref{sectionK4limits}) 
we solved the differential equations (\ref{DEK4}) for $\vec{f}$ to order $\eps^6$, i.e. weight six,
which is the typical weight required for three-loop computations.
The explicit results are given in attached text files.

Here, as an example and an application of these results, we will define the integral
\begin{align}\label{defK0}
K^{(0)}(x,\eps)=e^{3 \eps \gamma_{E}}  (-s)^{-3\eps}(1-4\eps)(1-5\eps)\eps^4 K_{1,1,1,1,1,1}(x,\eps)\,,
\end{align}
which is of special interest, as discussed in the introduction.
It is related to the basis above via
\begin{align}\label{defK0dec}
K^{(0)}(x,\eps) = \frac{1}{12} (3 f_1 - 3 f_2 + 3 f_3 + 11 f_5 - 11 f_6 + 11 f_7 - f_8 +
   f_9 - f_{10}) \,.
   \end{align}
  As a consequence, it has the same uniform weight properties as the $f_{i}$. 
The first terms of its expansion in $\eps$ are given by
\bea \label{resultK0maintext}
K^{(0)}(x,\eps)=&&
 2 \zeta _3 \epsilon ^3  \\ &&  \hspace*{0mm} 
 +\epsilon ^4 \biggl[ 3 i \pi  \zeta _3+\frac{3 \pi ^4}{20}
 +2 i \pi  H_{-3}(x)+\frac{1}{2} \pi ^2 H_{-2}(x)
-\frac{1}{2} i \pi ^3 H_{-1}(x) 
-3 H_{-1}(x) \zeta _3 \nn \\ &&  \hspace*{0mm} 
-2 H_{-3,-1}(x)+H_{-2,-2}(x)-i \pi  H_{-2,0}(x)+H_{-1,-3}(x)
-\pi ^2 H_{-1,-1}(x)\nn \\ &&  \hspace*{0mm} 
+\frac{1}{2} \pi ^2 H_{-1,0}(x)+H_{-2,-1,0}(x)
+H_{-1,-2,0}(x)-i \pi  H_{-1,0,0}(x)-2 H_{-1,-1,0,0}(x)\biggr] 
\nn \\ &&  \hspace*{0mm} 
+{\mathcal O}(\epsilon^5) \nn  \,.
\eea 
The terms of order $\eps^5$ and $\eps^6$ are presented in the appendix, for completeness.
Note that in eq. (\ref{resultK0maintext}), we have chosen to represent the answer in such a way
that the branch cuts of all functions involved lie on the negative real axis.\footnote{A word of caution is in order here. As was mentioned at the beginning of section \ref{sectionK4onshell},  the analytic continuation of $f(x)$ is {\it not} simply obtained by replacing
$x \to x-i0$. In particular, terms like $\log(1+x+i0)$ spoil this simple picture. However, as long as $x>-1$, this naive replacement is valid.
Therefore, the identities involving $x\to 1/x$ can be safely used in our results for $x>0$, while the identities that map $x \to -1-x$ can be used for $-1<x<0$. \label{footnote_caution}}

Finally, let us mention that integral $K_{0}$ of eq. (\ref{defK0}) is completely crossing symmetric (see subsection \ref{section:crossing}),  i.e.
\begin{align}
K_{0}(x) =& K_{0}(-1-x) \,, \label{crosssingK0a} \\ 
K_{0}(x) =&  x^{-3 \eps} K_{0}(1/x)  \label{crossingK0b} \,.
\end{align}
As was discussed in subsection \ref{section:crossing}, for $x<0$ eq. (\ref{crossingK0b}) is to be replaced by
\begin{align}
 K_{0}(x) =  \left((x+i 0)^{-3 \eps} K_{0}(1/x)\right)^* \,,\qquad\qquad x<0 \,.
\end{align}
One may explicitly verify eq. (\ref{crosssingK0a}) for $-1<x<1$ (cf. footnote \ref{footnote_caution}) and (\ref{crossingK0b}) for $x>0$
and this is a non-trivial test of our result (\ref{resultK0maintext}).
We have done so using the convenient Mathematica implementation of harmonic polylogarithms \cite{Maitre:2005uu}.

We stress that, as a consequence of the form of the differential eqs. (\ref{DEK4}), all integrals
are pure functions of uniform weight\footnote{Strictly speaking, the boundary constants could invalidate this conclusion. 
Here one can easily see that the bubble integrals $f_1, f_{3}, f_{5}$ are of uniform weight. Moreover, 
the crossing relations preserve the weight.
}.

\subsection{Further analytic and numerical checks}

The terms up to weight three are in agreement with analytical results which we obtained with the MB representation 
presented in Appendix~B.
 
We have checked our results for the master integrals numerically by {\tt FIESTA}
\cite{Smirnov:2008py,Smirnov:2009pb,FIESTA3}.
To do this we evaluated with {\tt FIESTA} the canonical master integrals because
evaluating integrals with an index equal to 2 is preferable to evaluating integrals with an index equal to $-1$.
Then the numerical results for the elements of our uniformly transcendental basis could be obtained because
we have an IBP reduction at hand.
The agreement between our analytical and numerical results was achieved at least at the level of three digits
in the $\eps$ expansion up to weight six.
 
\section{Evaluating $K_{4}$ integrals with one leg off-shell} 
\label{sectionK4offshell}

\subsection{The choice of master integrals and differential equations}

Here we study the same integral class as in the previous section, but this time with one off-shell leg, $p_{4}^2 \neq 0$. 
We will use the same notation as before, hoping that this will not lead to confusion.
Let us use the following independent variables, $s= (p_1 + p_2)^2$, $t= (p_1 + p_3)^2$ as before, and $u=(p_2+p_3)^2$. 
They are related to $p_{4}^2$ via $s+t+u = p_{4}^2$.
The integrals we consider now are defined by the same formulae (\ref{diC}) and (\ref{K4ints}) as in the case $p_{4}^2= 0$.
It is instructive to look at the $\alpha$ representation, see eq. (\ref{alpha-d-mod}),
where the polynomial $\cal U$ is given by eq. (\ref{defUK4}), and $\cal W$ is given by
\begin{align}
{\cal W}= &(-s) \al_2 \al_3 \left(\al_1 \al_4+\al_6 \al_4+\al_1 \al_6+\al_5 \al_6\right) +\nonumber \\
&(-t) \al_3 \al_6 \left(\al_1 \al_2+\al_4 \al_2+\al_5 \al_2+\al_4 \al_5\right) +\nonumber\\
&(-u) \al_2 \al_6  \left(\al_1 \al_3+\al_4 \al_3+\al_5 \al_3+\al_1 \al_5\right) \label{defW}\,.
\end{align}
Note that this agrees with eq. (\ref{defWK4}) for $u\to -s-t$. 
We can make the following useful observation.
Since the $\al_{i} \ge0$, we see that ${\cal W}$ eq. (\ref{defW}) is positive for  $s<0,t<0,u<0$.
As a consequence, the Feynman integrals are real in this kinematical region. 
This is a first advantage of having introduced an extra scale, since such a `Euclidean region' does not exist for $p_{4}^2=0$. 
Moreover, the same fact allows us to absorb the $+i0$ prescription into the definition
of $s,t,u$, by giving them a small positive imaginary part. We will leave this small
imaginary part implicit in the formulas below.
Finally, we can always go to dimensionless functions that only depend on two dimensionless variables, which we choose to be 
\begin{align}
x = t/s\,,\quad y = u/s \,.
\end{align}
Realizing this by setting $s=-1$, this means that $x$ and $y$ have a small negative imaginary part.

We will proceed in the same manner as in the previous section.
Solving the IBP identities one finds that there are $16$ master integrals. We choose them as follows,
\begin{align}\label{K4p4rescaling}
\vec{f} = e^{3 \eps \gamma_{E}}  (-s)^{-3\eps}  \vec{g}\,, 
\end{align}
with 
\begin{align}\label{defK4p4basis}
g_1=& \eps ^3 t \bar{K}_{0, 0, 1, 2, 2, 2}  , \qquad
g_2=  \eps ^4 (p_4^2 - t) \bar{K}_{0, 1, 2, 2, 1, 1} , \\
g_3=& \eps ^3 (p_4^2 - s - t) \bar{K}_{1, 2, 0, 0, 2, 2}  ,\qquad
g_4= \eps ^4 (s + t) \bar{K}_{2, 2, 1, 0, 1, 1} ,\\
   g_5=& \eps ^3 s \bar{K}_{1, 2, 2, 2, 0, 0}  ,\qquad
   g_6= \eps ^4 (p_4^2 - s) \bar{K}_{2, 2, 1, 1, 0, 1} ,\\
   g_7=& -\eps ^4 (s + t) (\hat{\bar{K}}_{1, 2, 1, 1, 2, 1,-1} 
+  \eps  \bar{K}_{1, 2, 1, 1, 1, 1}) ,\qquad
   g_8= \eps ^5 t \bar{K}_{2, 1, 1, 1, 1, 1} ,\\
   g_9= & \eps ^5 (s + t) \bar{K}_{1, 1, 2, 1, 1, 1} ,\qquad
   g_{10}= \eps ^5 s \bar{K}_{1, 1, 1, 1, 2, 1}  ,  \\
    g_{11}=&\eps ^5 (p_4^2 - t) \bar{K}_{1, 2, 1, 1, 1, 1}   ,\qquad
   g_{12}= \eps ^5 (p_4^2 - s - t) \bar{K}_{1, 1, 1, 2, 1, 1} ,  \\
    g_{13}= &\eps ^5 (p_4^2 - s) \bar{K}_{1, 1, 1, 1, 1, 2}  ,\qquad
   g_{14}= \eps ^4 s (p_4^2 - s - t) \bar{K}_{2, 2, 1, 1, 1, 1} ,  \\
    g_{15}= &\eps ^4 s t \bar{K}_{1, 1, 2, 2, 1, 1} ,\qquad
   g_{16}= \eps ^4 t (p_4^2 - s - t) \bar{K}_{1, 1, 1, 1, 2, 2} ,  
\end{align}
where we use definitions (\ref{K4ints}) and (\ref{K4ints1}) and the bar denotes integrals with $p_4^2\neq 0$.

Just as in the on-shell case, we can relate the master integrals above
to the integral with unit powers of the propagators,
\begin{align}
\bar{K}^{(0)} = e^{3 \eps \gamma_{E}}  (-s)^{-3\eps}   \eps^4 (1 - 4 \eps) (1 - 5 \eps) \bar{K}_{1,1,1,1,1,1} \,.
\end{align}
We have
\begin{align}
\bar{K}^{(0)} =&  \frac{1}{48} (
5 f_1+28 f_2+5 f_3+28 f_4+5 f_5+28 f_6   +156 f_8-20 f_9
  \nonumber\\
   &\hspace{0.4 cm}+156 f_{10} -20 f_{11}+156 f_{12}-20
   f_{13}  -8 f_{14}-8 f_{15}-8 f_{16} ) \,.
\end{align}

Dealing with several variables does not modify the approach.
We derive partial differential equations in $s, t$ and $u$.
They are conveniently written in a differential form,
\begin{align}\label{DEK4p4}
d \,\vec{f}(s,t,u;\eps) =& \eps\,  d \tilde{A}(s,t,u) \,  \vec{f}(s,t,u;\eps)\,,
\end{align}
with 
\begin{align}\label{AtildeK4p4}
\tilde{A}(s,t,u) =&  \big[ A_1 \log(-s)  +  A_2 \log(-t)  +  A_3 \log(-u) +  A_4 \log(-s-t-u)   \nonumber \\ & \quad + A_{5} \log(-s-t) + A_{6} \log(-s-u) + A_{7} \log(-t-u)  \big]\,.
\end{align}
and where the $A_{i}$ are {\it constant} $16 \times 16$ matrices. 
In fact, since $\vec{f}$ is dimensionless, this is really a two-variable problem,
i.e. $\sum_{i=1}^{7} A_{i} = 0_{16 \times 16}$.
We have preferred, however, to write it in this more symmetric form.

The alphabet of differential forms we obtain is the same as the one occurring at the previous loop order. (We have verified that the same form of the equations (\ref{DEK4p4}) and (\ref{AtildeK4p4}) holds at two loops, by choosing a slightly different basis compared to \cite{Gehrmann:2000zt,Gehrmann:2001ck}.)

Leaving the issue of the boundary conditions aside for the moment (this will be dealt with in the next subsection), eq. (\ref{DEK4p4}) allows us to solve for $\vec{f}$ to any desired order in the $\eps$ expansion. It is clear that each term in the answer will be given by iterated integrals over the differential one-forms shown above. This class of functions forms a subset of multiple polylogarithms and was studies in \cite{Gehrmann:2000zt,Gehrmann:2001ck}.

More generally, we can write the solution to eq. (\ref{DEK4p4}) in the beautiful language of
Chen iterated integrals. Let $M$ be a (in general complex) manifold describing the kinematical data, in this case $s,t,u$. Each element of the matrix $d\tilde{A}$ is a one-form on this manifold.
The integration contour is then a path on this manifold. We can parametrize it by defining a map $\gamma : [0,1] \rightarrow M$.
Denoting the pull-back of the form $d\tilde{A}$ to the interval $[0,1]$ by $A(\tau) d \tau$, 
a line integral is then given by
\begin{align}
\int_{\gamma} d\tilde{A} = \int_0^{1} A(\tau_1) d \tau_1 \,.
\end{align}
The iterated integrals we are interested in are then defined as
\begin{align}
\int_\gamma \underbrace{d\tilde{A} \ldots d\tilde{A}}_{n} = \int_{0\le \tau_1 \le \ldots \le \tau_n\le 1} A(\tau_1) 
d \tau_1 \ldots A(\tau_n)d \tau_n \,.
\end{align}
Using these iterated integrals, we can write down the general solution to eq. (\ref{DEK4p4}).
It is given by
\begin{align}\label{solutionK4p4}
\vec{f}(s,t,u;\eps) = \mathcal{P} \exp\left[ \eps \int_{\gamma} d\tilde{A} \right] \vec{h}(\eps)\,,
\end{align}
where $\vec{h}(\eps)$ represents the boundary condition.
Expanding the exponential in eq. (\ref{solutionK4p4}) perturbatively in $\eps$, one obtains
at order $\eps^k$ (linear combinations of) $k$-fold iterated intervals.
The latter are homotopy invariant line integrals, with $\gamma$ connecting the base point $(s_0, t_0, u_0)$ to the argument of the function, $(s,t,u)$. 
Upon choosing a specific contour of integration, one can recover expressions in terms of multiple polylogarithms.
In the next section, we will provide the information for determining the boundary constants.

\subsection{Determining the boundary conditions}
\label{section_K4p4boundary}

There are several boundary conditions that we can use, as we discuss presently:
\begin{itemize}
\item Elementary integrals: integrals $f_{1}, f_{3},f_{5}$ are trivial bubble-type integrals 
that can be expressed in terms of $\Gamma$ functions.
\item Branch cut structure: for massless integrals, we expect branch cuts to start only at positions $p_{i}^2 = 0, (p_{i} + p_{j})^2=0$, etc. 
Inspecting the terms on the r.h.s. of eq. (\ref{DEK4p4}), we see that only the logarithms on the first line have arguments of this form. Imposing the absence of branch cuts coming from functions like $\log(-s-t)$ then imposes constraints on the answer.
We have verified this expectation using the computer code {\tt asy.m} mentioned earlier.
\item Asymptotic limit and UV behavior: just as in the previous section, we can determine some of the coefficients in the asymptotic expansion by requiring the absence of UV divergences in the basis $\vec{f}$.
For example, in this way it can be seen that this implies 
that $f_{8} \propto  K_{2,1,1,1,1,1} \to 0$ as $t\to0$.
\item Symmetry relations: some integrals are symmetric under exchange of $p_{1}$ and $p_{3}$, e.g. $(-s)^{3 \eps} f_{9}(s,t,u;\eps) = (-t)^{3 \eps} f_{9}(t,s,u;\eps)$. Other integrals are mapped into each other under this exchange, e.g. $(-s)^{3 \eps} f_{8}(s,t,u;\eps) = (-t)^{3 \eps} f_{10}(t,s,u;\eps)$.
\item Simple limits: In general, the limits at the singular points of the DE do not commute with the $\eps$ expansion. However, for some integrals the situation is simpler. 
For example, for integral $K^{(0)}$ one expects soft limits such as $p_{1}\to 0$ to commute with the $\eps$ expansion, since they do not change the divergence structure of the integral. (The integral is IR finite, and stays IR finite in the limit. The UV divergences are unchanged by the limit.) At $p_{1} = 0$, 
however, $K_{1,1,1,1,1,1}$ becomes a known planar form-factor integral, and we can use its value as boundary 
condition.\footnote{Note that the form-factor integral can itself be determined by bootstrap arguments \cite{Henn:2013tua}.}
\end{itemize}
We have found the above requirements to be sufficient to determine all boundary constants for all $16$ integrals, order by order in $\eps$. In fact, one can see that the first three elements on the list above are sufficient to fix all integration constants.
We stress that these conditions do not require any integrations and can be implemented in an algebraic way.

 \subsection{Analytic solution and on-shell limit}
Here we present the analytic solution for the first orders in the $\eps$ expansion of the integrals.
As discussed above, it can be written as (\ref{solutionK4p4}), with the boundary conditions following from the considerations in the previous paragraph. 
In (\ref{solutionK4p4}), one has the freedom of choosing a base point for the iterated integral.
One reasonable choice would be $s=t=u=-1$, since this stays away from all potentially singular
surfaces. However, this leads to rather awkward integrations, such as $\log 3$ at weight one.
In the literature, results for the alphabet (\ref{AtildeK4p4}) are usually represented in terms
of so-called two-dimensional harmonic polylogarithms, a subset of Goncharov polylogarithms (GPL).

GPL are defined as follows.
\begin{align}
G(a_1,\ldots, a_n ; z) = \int_0^z \frac{dt}{t-a_{1}} G(a_{2}, \ldots ,a_{n}; t) \,,
\end{align}
with 
\begin{align}
G(a_1 ;z) = \int_0^z \frac{dt}{t-a_{1}}  \,, \qquad a_{1} \neq 0\,.
\end{align}
For $a_{i}=0$, we have $G(\vec{0}_{n};x) = 1/n! \log^n(x)$.
The total differential of a general Goncharov polylogarithm is 
\begin{align}\label{differentialG}
d G(a_1 , \ldots, a_{n};z) =&\; G(\hat{a}_{1}, a_{2}, \ldots a_{n}; z) \, d \log\frac{z-a_1}{a_1 - a_2} \nonumber \\
& + G(a_{1}, \hat{a}_{2}, a_{3} , \ldots, a_{n}; z) \,  d \log \frac{a_1 - a_2 }{a_2 - a_3} + \ldots +  \nonumber \\
& + G(a_{1}, \ldots, a_{n-1}, \hat{a}_{n}; z) \, d \log \frac{a_{n-1}-a_{n}}{a_{n}} \,,
\end{align}
where $\hat{a}$ means that this element is omitted.

The subset of two-dimensional harmonic polylogarithms is obtained by specifying labels to be from the set $\{0,-1,-1-y,-y\}$ and argument $z=x$. 
One can easily convince oneself that this set of functions, together with HPL of argument $y$,
is sufficient in order to represent the solution (\ref{solutionK4p4}).  
It essentially corresponds to choosing $s=-1,t=0,u=1$ as base point (after separating the logarithmic divergences as $t \to 0$.)

Here we followed a slightly different approach, by first integrating the differential equation in $x$, and then in $y$. The procedure is almost the same as in \cite{Henn:2013woa}.

The boundary constants are determined from the conditions discussed in section \ref{section_K4p4boundary}. When approaching the various limits discussed there,
one sometimes encounters spurious divergences of the Goncharov polylogarithms.
This is a slight disadvantage of having gone from the language of Chen iterated
integrals to the latter. Such divergences can be extracted before taking limits by
using the shuffle product formula of Goncharov polylogarithms,
\begin{align}
G(a_1, \ldots, a_n ;z) G(b_1, \ldots, b_m;z ) =& G(\vec{a} ;z) G(\vec{b}; z) = \sum_{\vec{c} \in \vec{a} \,\uplus \,  \vec{b}}  G(\vec{c} ;z) \,,
\end{align}
where $ \vec{a} \,\uplus \,  \vec{b}$ is the shuffle product of two ordered sets, i.e. all combined sets where the relative order of the elements of $\vec{a}$ and $\vec{b}$ is preserved.

In this way, the complete solution can be obtained algorithmically.
We give an example for illustration. For integral $f_{8}$ up to order $\eps^2$, we have
\begin{align}
f_{8} =&  \frac{1}{4} \eps^2 \Big[ G_{-1,-1-y}(x)+G_{-y,-1-y}(x)+ G_{-1}(x)
   H_{-1}(y) +H_{-1}(y) G_{-y}(x)\nonumber \\ & \hspace{0.5cm} -G_{-1}(x) H_0(y) \Big] + \cO(\eps^3) \,,
   \end{align}
and similarly for the other integrals.
Of course, the results up to weight two can easily be rewritten in terms of logarithms and dilogarithms,
and probably in a more compact way.
We prefer to use the language of GPL and HPL because it is valid at any order in $\eps$.

Finally, we wish to outline how to recover the on-shell case discussed in section \ref{sectionK4onshell}. In this way, one sees that the boundary constants for those integrals
also follow from the general considerations made here. 
The limit $p_{4}^2 \to 0$ is governed by the term $\eps A_{4} \log(-s-t-u)$ in the differential equation.
Analyzing $\eps A_{4}$, we find that  it has two possible eigenvalues $-2 \eps$ and $0$.
One can then proceed as explained in section \ref{sectionasymptoticDE1} in order to 
resolve the order of limits ambiguities, and obtain results for $K_{4}$ on-shell.

This completes our discussion of the $K_{4}$ integrals. In summary, we have seen that the latter
are completely determined by the differential equations discussed here, and that the boundary constants follow from simple physical considerations. In our setup, it is clear that the results are pure functions of uniform weight in the $\eps$ expansion.
We have also outlined how to recover results in the on-shell case from this setup.
Note that in this approach, in contrast to section \ref{sectionK4limits}, no integrals have to be calculated in order to determine the boundary conditions.

\section{Discussion and outlook}
\label{sec:discussion}

In this paper we observed that the differential equations for master integrals can be 
used to infer the structure of asymptotic expansions of the master integrals. 
We applied this information to the computation of single-scale and
non-planar integrals. Although we mainly had in mind to give examples
showing the scope of this method, many of the results derived in this paper are new,
and the integrals computed are highly non-trivial.
In this paper, we mainly focused on using the information on singular limits in order to provide simple
boundary conditions for the DE. Of course, one can also go the other way. We expect that this will have many applications for the analysis of physically  interesting limits.

In order to use the method of DE for single-scale integrals, we first generalized the problem by introducing an extra scale.
On the one hand, this leads to an increase of the number of master integrals needed, from 14 to 39.
On the other hand, it allows to use the powerful DE technique. In fact, once the basis of master integrals
is chosen appropriately \cite{Henn:2013pwa}, the number of integrals does not play an important role in the structure of the equations.
Moreover, the additional scale gives access to new limits where the boundary constants can be determined easily.
In this way, we solved the more general two-scale problem, using only algebraic means.
Finally, the knowledge of the precise scaling behavior of the Feynman integrals, also inferred from the
differential equations, allowed us to relate the two-scale problem to the one-scale problem we started with.

The procedure leading from the Feynman integrals to the set of differential equations for master integrals
is entirely algebraic. The differential equations, especially when written in the simple form of \cite{Henn:2013pwa},
make it clear which class of special functions is needed to describe the Feynman integrals, to all orders in $\eps$.
In particular, this also determines what types of transcendental constants can appear at special values of these functions.
These periods of Feynman integrals are heavily studied in the mathematical literature. The approach proposed here, which consists of solving a more general problem via differential equations  and 
then to obtain the periods as a corollary, has also been used in the mathematical literature \cite{Brown:2012ia}. Feynman integrals depending on a parameter, called graphical functions there, were computed there with the help of the second-order differential equations of \cite{Drummond:2006rz,Drummond:2010cz}. The desired
periods where then obtained at special values of those functions, and this was used to prove a conjecture made in ref. \cite{Broadhurst:1995km}.

The second class of Feynman integrals computed in this paper is a family of non-planar on-shell
four-point functions. We consider integrals corresponding to the graph $K_{4}$.
They were found not to be linearly reducible in the framework of ref. \cite{Bogner:2013tia}.
Here we studied them in the context of DE and found that the DE have the same form as 
in the previously studied planar case \cite{Henn:2013tua}, and in particular, lead to the same class of multiple polylogarithms and transcendental constants.

We did two calculations for these integrals, the first one, in section \ref{sectionK4onshell} being a direct one. In order to determine the integration constants, we used information from the asymptotic expansion of the integrals. This was possible thanks to the control the DE give over such expansions.
In section  \ref{sectionK4offshell}, we performed the calculation for the same integrals with one external leg off-shell. As in the case of the form-factor integrals, the number of master integrals
increased, here from $10$ to $16$. On the other hand, this allowed us to fix the integration constants in a clear and simple way. We outlined how the results for the on-shell integrals can be recovered.
This was done mainly as a proof of principle that no integrals have to be performed in order to
find the integration constants.

The method and results presented here for the $K_{4}$ integrals strongly suggest to us that the planar
results of \cite{Henn:2013tua} can be carried over to the non-planar case. The results presented here and in \cite{Henn:2013tua} already allow the computation of non-planar scattering amplitudes in 
$\phi^4$ models.
Completing the calculation for all non-planar master integrals will allow the computation of
non-planar scattering amplitudes in super Yang-Mills and supergravity theories
that are currently only known at the integrand level \cite{Bern:2007hh}.
This will give valuable insights into the generic structure of infrared divergences in gauge and gravity theories.
The methods developed in the present paper should be extremely helpful in determining the boundary
constants for the required non-planar integrals.

\vspace{0.2 cm}
{\em Acknowledgments.}
%
J.H. thanks the Simons Center for hospitality during part of this work, and the organizers of RADCOR 2013 for their invitation.
 J.H. is supported in part by the DOE grand DE-SC0009988 and by the Marvin L. Goldberger fund.
The work of V.S. was supported by the Alexander von Humboldt Foundation (Humboldt Forschungspreis).

\appendix

\section{Evaluating  $K_4$ integrals by Mellin-Barnes representation}

Since we are dealing with a non-planar graph,
the loop-by-loop strategy of deriving MB representations is not optimal so that one is forced
to derive them by hand separating various terms in the basic functions of alpha parameters at
the cost of introducing MB integrations.
We have derived the following eight-fold MB representation for (\ref{K4ints}):
\bea
K_{a_1,\ldots,a_6}=
\frac{1}{\Gm(8 - a - 4 \eps ) \prod_{i}\Gm(a_i)}
\frac{1}{(2\pi i)^8}
\int_{-i\infty}^{+i\infty}\ldots \int_{-i\infty}^{+i\infty}
\frac{(-s-i 0)^{z_2} (-t-i 0)^{z_1}}{(s+t-i 0)^{a + 3 \eps -6  + z_1 + z_2}}
\nn && \\ &&  \hspace*{-130mm}
\times
\frac{\Gm(-z_1) \Gm(-z_2) 
\Gm(a + 3 \eps -6  + z_1 + z_2) 
\Gm(a_{1345} + 2 \eps-4  + z_1 + z_2 - z_3) \Gm(z_6-z_5 ) }
{\Gm(a_{135} - a_2  + \eps -2  + z_1 - z_4) \Gm(6 - a_{13456} - 3 \eps  - z_1 + z_5) }
\nn \\ &&  \hspace*{-130mm}\times
\frac{ \Gm(8 - a_{1233456} - 4 \eps  - z_1 - z_2 + z_3) 
\Gm(2 - a_{24} - \eps  - z_2 + z_3 - z_4) 
  \Gm(z_3 - z_4 + z_6)}
{\Gm(10 - a_{12334556} - 5 \eps  - z_1 + z_4)}
\nn \\ &&  \hspace*{-130mm}\times
\Gm(6 - a_{23456} - 3 \eps  - z_1 + z_5) 
\Gm(6 - a_{13456} - 3 \eps  - z_1 + z_3 + z_5 - z_6)  \Gm( z_6-z_3 ) 
\nn \\ &&  \hspace*{-130mm}\times
\Gm(6 - a_{13456} - 3 \eps  - z_1 - z_3 + z_4 + z_5 - z_6) 
\Gm(-z_6)\Gm(2 - a_5 - \eps  + z_2 - z_3 + z_4)
\nn \\ &&  \hspace*{-130mm}\times
\Gm(4 - a_{135} - 2 \eps  + z_4 - z_6)
\Gm(a_{1334556} + 4 \eps -8  + 2 z_1 - z_4 - z_5 + z_6)
\prod_{j=1}^8 d z_j
\; ,
\label{8fMB}
\eea  
 where $a_{12334556}=a_1+a_2+2a_3+a_4+2a_5+a_6$, etc. and $a=a_1+\ldots+a_6$.

However, at concrete integer indices, there is usually the possibility to take two
integrations by means of the first Barnes lemma. In particular, we obtain
\bea
K_{1,\ldots,1}=
\frac{1}{\Gm(2 - 4 \eps ) \Gm(1 - 2 \eps )^2}
\frac{1}{(2\pi i)^6}
\int_{-i\infty}^{+i\infty}\ldots \int_{-i\infty}^{+i\infty}
\frac{(-s-i 0)^{z_2} (-t-i 0)^{z_1}}{(s+t-i 0)^{3 \eps+ z_1 + z_2}}\Gm(-z_1) \Gm(-z_2) 
\nn && \\ &&  \hspace*{-150mm}
\times
\frac{\Gm(3 \eps  + z_1 + z_2) 
\Gm(\eps  + z_1 - z_3) \Gm(2 \eps  + z_1 + z_2 - z_3)\Gm(z_3 - z_4)\Gm(\eps  + z_1 + z_3 - z_4) }
{\Gm(\eps  + z_1 - z_4) \Gm(2 - 5 \eps  - z_1 + z_4)}
\nn \\ &&  \hspace*{-150mm}\times
\Gm(1 - 2 \eps  + z_3) \Gm(1 - 3 \eps  - z_1 + z_3)  \Gm(-z_3)\Gm(1 - 4 \eps  - z_1 - z_2 + z_3) 
 \Gm(-\eps  - z_2 + z_3 - z_4)
\nn \\ &&  \hspace*{-150mm}\times
  \Gm(
    1 - 2 \eps  - z_3 + z_4) \Gm(1 - 3 \eps  - z_1 - z_3 + z_4) \Gm(
    1 - \eps  + z_2 - z_3 + z_4))
 \prod_{j=1}^6 d z_j
\; .
\label{6fMB}
\eea  

To evaluate this and other above mentioned master integrals one can apply public computer codes
\cite{Czakon:2005rk,Smirnov:2009up,heptools1,heptools2}.
This straightforward procedure works well only up to weight three in the $\eps$-expansion.

\section{Explicit results for $K^{(0)}$ up to weight six}
  
Here we present analytical results for the  integral (\ref{defK0})
which is expresed in terms of the master integrals
discussed in the main text via  eq.~(\ref{defK0dec})
and therefore   
   has the same uniform weight properties as the $f_{i}$. 
We have 
\bea
K^{(0)}(x,\eps)=
 2 \zeta _3 \epsilon ^3+\epsilon ^4 \biggl(3 i \pi  \zeta _3+\frac{3 \pi ^4}{20}
 +2 i \pi  H_{-3}+\frac{1}{2} \pi ^2 H_{-2}
-\frac{1}{2} i \pi ^3 H_{-1}-3 H_{-1} \zeta _3
&& \nn \\ &&  \hspace*{-135mm} 
-2 H_{-3,-1}+H_{-2,-2}-i \pi  H_{-2,0}+H_{-1,-3}-\pi ^2 H_{-1,-1}
+\frac{1}{2} \pi ^2 H_{-1,0}+H_{-2,-1,0}
\nn \\ &&  \hspace*{-135mm} 
+H_{-1,-2,0}-i \pi  H_{-1,0,0}-2 H_{-1,-1,0,0}\biggr)   
\nn \\ &&  \hspace*{-135mm}  
+\epsilon ^5 \biggl(\frac{9 i \pi ^5}{40}-5 \pi ^2 \zeta _3+50 \zeta _5+4 i \pi  H_{-4}-\frac{7}{2} \pi ^2 H_{-3}+\frac{3}{2} i \pi ^3 H_{-2}
-15 i \pi  H_{-1} \zeta _3-\frac{1}{10} \pi ^4 H_{-1}
\nn \\ &&  \hspace*{-135mm}
-4 H_{-4,-1}-H_{-3,-2}-12 i \pi  H_{-3,-1}+i \pi  H_{-3,0}-4 H_{-2,-3}-6 i \pi  H_{-2,-2}+\frac{5}{2} \pi ^2 H_{-2,-1}
\nn \\ &&  \hspace*{-135mm} 
-\frac{1}{2} \pi ^2 H_{-2,0}-6 H_{-1,-4}-8 i \pi  H_{-1,-3}+\frac{5}{2} \pi ^2 H_{-1,-2}+\frac{1}{2} i \pi ^3 H_{-1,-1}-3 H_{-1,-1} \zeta _3
\nn \\ &&  \hspace*{-135mm}
+\frac{3}{2} i \pi ^3 H_{-1,0}
+12 H_{-3,-1,-1}-H_{-3,-1,0}+6 H_{-2,-2,-1}-4 H_{-2,-2,0}-6 H_{-2,-1,-2}
\nn \\ &&  \hspace*{-135mm} 
+6 i \pi  H_{-2,-1,0}+4 i \pi  H_{-2,0,0}+8 H_{-1,-3,-1}-6 H_{-1,-3,0}-4 H_{-1,-2,-2}+4 i \pi  H_{-1,-2,0}
\nn \\ &&  \hspace*{-135mm} 
-H_{-1,-1,-3}
-2 \pi ^2 H_{-1,-1,-1}+\frac{5}{2} \pi ^2 H_{-1,-1,0}-\frac{3}{2} \pi ^2 H_{-1,0,0}-6 H_{-2,-1,-1,0}
\nn \\ &&  \hspace*{-135mm} 
+8 H_{-2,-1,0,0}-4 H_{-1,-2,-1,0}+6 H_{-1,-2,0,0}-H_{-1,-1,-2,0}+i \pi  H_{-1,-1,0,0}+6 i \pi  H_{-1,0,0,0}
\nn \\ &&  \hspace*{-135mm} 
-4 H_{-1,-1,-1,0,0}+12 H_{-1,-1,0,0,0}\biggr)
\nn \\ &&  \hspace*{-135mm}
+\epsilon ^6 \biggl(-\frac{21}{4} i \pi ^3 \zeta _3+75 i \pi  \zeta _5-\frac{31 \pi ^6}{126}-70 \zeta _3^2+6 i \pi  H_{-5}
-\frac{15}{2} \pi ^2 H_{-4}-\frac{11}{3} i \pi ^3 H_{-3}+34 i \pi  H_{-2} \zeta _3
\nn \\ &&  \hspace*{-135mm}
-\frac{37}{24} \pi ^4 H_{-2}
-\frac{19}{30} i \pi ^5 H_{-1}+\frac{73}{4} \pi ^2 H_{-1} \zeta _3-75 H_{-1} \zeta _5-6 H_{-5,-1}-3 H_{-4,-2}-24 i \pi  H_{-4,-1}
\nn \\ &&  \hspace*{-135mm}
+3 i \pi  H_{-4,0}+2 H_{-3,-3}-10 i \pi  H_{-3,-2}
+19 \pi ^2 H_{-3,-1}-\frac{1}{2} \pi ^2 H_{-3,0}+15 H_{-2,-4}-6 i \pi  H_{-2,-3}
\nn \\ &&  \hspace*{-135mm}
+\frac{13}{4} \pi ^2 H_{-2,-2}-\frac{35}{6} i \pi ^3 H_{-2,-1}+32 H_{-2,-1} \zeta _3
-\frac{11}{12} i \pi ^3 H_{-2,0}+27 H_{-1,-5}+\frac{1}{4} \pi ^2 H_{-1,-3}
\nn \\ &&  \hspace*{-135mm}
-\frac{23}{6} i \pi ^3 H_{-1,-2}+20 H_{-1,-2} \zeta _3+9 i \pi  H_{-1,-1} \zeta _3
-\frac{2}{15} \pi ^4 H_{-1,-1}+10 i \pi  H_{-1,0} \zeta _3-\frac{7}{24} \pi ^4 H_{-1,0}
\nn \\ &&  \hspace*{-135mm}
+24 H_{-4,-1,-1}-3 H_{-4,-1,0}+10 H_{-3,-2,-1}+2 H_{-3,-2,0}+6 H_{-3,-1,-2}+54 i \pi  H_{-3,-1,-1}
\nn \\ &&  \hspace*{-135mm}
-6 i \pi  H_{-3,-1,0}-2 i \pi  H_{-3,0,0}+6 H_{-2,-3,-1}+15 H_{-2,-3,0}+14 H_{-2,-2,-2}+45 i \pi  H_{-2,-2,-1}
\nn \\ &&  \hspace*{-135mm}
-14 i \pi  H_{-2,-2,0}
+20 H_{-2,-1,-3}+25 i \pi  H_{-2,-1,-2}+\frac{9}{2} \pi ^2 H_{-2,-1,-1}-\frac{63}{4} \pi ^2 H_{-2,-1,0}
\nn \\ &&  \hspace*{-135mm}
+\frac{3}{2} \pi ^2 H_{-2,0,0}+27 H_{-1,-4,0}+20 H_{-1,-3,-2}+57 i \pi  H_{-1,-3,-1}-20 i \pi  H_{-1,-3,0}+14 H_{-1,-2,-3}
\nn \\ &&  \hspace*{-135mm}
+13 i \pi  H_{-1,-2,-2}+\frac{9}{2} \pi ^2 H_{-1,-2,-1}-\frac{51}{4} \pi ^2 H_{-1,-2,0}+6 H_{-1,-1,-4}+2 i \pi  H_{-1,-1,-3}
\nn \\ &&  \hspace*{-135mm}
+\frac{9}{2} \pi ^2 H_{-1,-1,-2}+\frac{3}{2} i \pi ^3 H_{-1,-1,-1}-3 H_{-1,-1,-1} \zeta _3
-\frac{11}{6} i \pi ^3 H_{-1,-1,0}+8 H_{-1,-1,0} \zeta _3
\nn \\ &&  \hspace*{-135mm}
-\frac{47}{12} i \pi ^3 H_{-1,0,0}-54 H_{-3,-1,-1,-1}+6 H_{-3,-1,-1,0}-2 H_{-3,-1,0,0}-45 H_{-2,-2,-1,-1}
\nn \\ &&  \hspace*{-135mm}
+14 H_{-2,-2,-1,0}-13 H_{-2,-2,0,0}
-25 H_{-2,-1,-2,-1}+20 H_{-2,-1,-2,0}+27 H_{-2,-1,-1,-2}
\nn \\ &&  \hspace*{-135mm}
-27 i \pi  H_{-2,-1,-1,0}
-20 i \pi  H_{-2,-1,0,0}-15 i \pi  H_{-2,0,0,0}-57 H_{-1,-3,-1,-1}+20 H_{-1,-3,-1,0}
\nn \\ &&  \hspace*{-135mm}
-25 H_{-1,-3,0,0}-13 H_{-1,-2,-2,-1}+14 H_{-1,-2,-2,0}+15 H_{-1,-2,-1,-2}-15 i \pi  H_{-1,-2,-1,0}
\nn \\ &&  \hspace*{-135mm}
-14 i \pi  H_{-1,-2,0,0}-2 H_{-1,-1,-3,-1}+6 H_{-1,-1,-3,0}+2 H_{-1,-1,-2,-2}-2 i \pi  H_{-1,-1,-2,0}
\nn \\ &&  \hspace*{-135mm}
-3 H_{-1,-1,-1,-3}-3 \pi ^2 H_{-1,-1,-1,-1}
+\frac{9}{2} \pi ^2 H_{-1,-1,-1,0}-7 \pi ^2 H_{-1,-1,0,0}+\frac{9}{2} \pi ^2 H_{-1,0,0,0}
\nn \\ &&  \hspace*{-135mm}
+27 H_{-2,-1,-1,-1,0}-57 H_{-2,-1,0,0,0}+15 H_{-1,-2,-1,-1,0}+6 H_{-1,-2,-1,0,0}
\nn \\ &&  \hspace*{-135mm}
-45 H_{-1,-2,0,0,0}+2 H_{-1,-1,-2,-1,0}+10 H_{-1,-1,-2,0,0}-3 H_{-1,-1,-1,-2,0}
\nn \\ &&  \hspace*{-135mm}
+3 i \pi  H_{-1,-1,-1,0,0}-6 i \pi  H_{-1,-1,0,0,0}-27 i \pi  H_{-1,0,0,0,0}-6 H_{-1,-1,-1,-1,0,0}
\nn \\ &&  \hspace*{-135mm}
+24 H_{-1,-1,-1,0,0,0}-54 H_{-1,-1,0,0,0,0}\biggr)   
   + \cO(\eps^7) \,.
\eea
Here the argument $x=t/s$ is omitted in all the HPL, for brevity.

\bibliographystyle{JHEP}

\bibliography{bibfile}

\providecommand{\href}[2]{#2}\begingroup\raggedright\begin{thebibliography}{10}

\bibitem{Kotikov:1990kg}
A.~V. Kotikov, {\it {Differential equations method: New technique for massive
  Feynman diagrams calculation}},  {\em Phys. Lett.} {\bf B254} (1991)
  158--164.

\bibitem{Kotikov:1991pm}
A.~V. Kotikov, {\it {Differential equation method: The Calculation of N point
  Feynman diagrams}},  {\em Phys. Lett.} {\bf B267} (1991) 123--127.

\bibitem{Remiddi:1997ny}
E.~Remiddi, {\it {Differential equations for Feynman graph amplitudes}},  {\em
  Nuovo Cim.} {\bf A110} (1997) 1435--1452,
  [\href{http://xxx.lanl.gov/abs/hep-th/9711188}{{\tt hep-th/9711188}}].

\bibitem{Gehrmann:1999as}
T.~Gehrmann and E.~Remiddi, {\it {Differential equations for two-loop
  four-point functions}},  {\em Nucl. Phys.} {\bf B580} (2000) 485--518,
  [\href{http://xxx.lanl.gov/abs/hep-ph/9912329}{{\tt hep-ph/9912329}}].

\bibitem{Gehrmann:2000zt}
T.~Gehrmann and E.~Remiddi, {\it {Two-Loop Master Integrals for $\gamma^* \to
  3$ Jets: The planar topologies}},  {\em Nucl. Phys.} {\bf B601} (2001)
  248--286, [\href{http://xxx.lanl.gov/abs/hep-ph/0008287}{{\tt
  hep-ph/0008287}}].

\bibitem{Gehrmann:2001ck}
T.~Gehrmann and E.~Remiddi, {\it {Two loop master integrals for $\gamma^* \to
  3$ jets: The Nonplanar topologies}},  {\em Nucl.Phys.} {\bf B601} (2001)
  287--317, [\href{http://xxx.lanl.gov/abs/hep-ph/0101124}{{\tt
  hep-ph/0101124}}].

\bibitem{Henn:2013pwa}
J.~M. Henn, {\it {Multiloop integrals in dimensional regularization made
  simple}},  {\em Phys.Rev.Lett.} {\bf 110} (2013) 251601,
  [\href{http://xxx.lanl.gov/abs/1304.1806}{{\tt arXiv:1304.1806}}].

\bibitem{Chen1997}
K.-T. Chen, {\it {Iterated path integrals}},  {\em Bull. Amer. Math. Soc.} {\bf
  83, Number 5} (1997) 831--879.

\bibitem{Goncharov:1998kja}
A.~B. Goncharov, {\it {Multiple polylogarithms, cyclotomy and modular
  complexes}},  {\em Math.Res.Lett.} {\bf 5} (1998) 497--516,
  [\href{http://xxx.lanl.gov/abs/1105.2076}{{\tt arXiv:1105.2076}}].

\bibitem{arXiv:math/0606419}
F.~Brown, {\it {Multiple zeta values and periods of moduli spaces
  $\mathfrak{M}_{0,n}$}},  \href{http://xxx.lanl.gov/abs/0606419}{{\tt
  0606419}}.

\bibitem{Henn:2013tua}
J.~M. Henn, A.~V. Smirnov, and V.~A. Smirnov, {\it {Analytic results for planar
  three-loop four-point integrals from a Knizhnik-Zamolodchikov equation}},
  {\em JHEP} {\bf 1307} (2013) 128,
  [\href{http://xxx.lanl.gov/abs/1306.2799}{{\tt arXiv:1306.2799}}].

\bibitem{Henn:2013woa}
J.~M. Henn and V.~A. Smirnov, {\it {Analytic results for two-loop master
  integrals for Bhabha scattering I}},  {\em JHEP} {\bf 1311} (2013) 041,
  [\href{http://xxx.lanl.gov/abs/1307.4083}{{\tt arXiv:1307.4083}}].

\bibitem{Gehrmann:2006wg}
T.~Gehrmann, G.~Heinrich, T.~Huber, and C.~Studerus, {\it {Master integrals for
  massless three-loop form-factors: One-loop and two-loop insertions}},  {\em
  Phys.Lett.} {\bf B640} (2006) 252--259,
  [\href{http://xxx.lanl.gov/abs/hep-ph/0607185}{{\tt hep-ph/0607185}}].

\bibitem{Heinrich:2007at}
G.~Heinrich, T.~Huber, and D.~Maitre, {\it {Master integrals for fermionic
  contributions to massless three-loop form-factors}},  {\em Phys.Lett.} {\bf
  B662} (2008) 344--352, [\href{http://xxx.lanl.gov/abs/0711.3590}{{\tt
  arXiv:0711.3590}}].

\bibitem{Baikov:2009bg}
P.~A. Baikov, K.~G. Chetyrkin, A.~V. Smirnov, V.~A. Smirnov, and
  M.~Steinhauser, {\it {Quark and gluon form factors to three loops}},  {\em
  Phys.Rev.Lett.} {\bf 102} (2009) 212002,
  [\href{http://xxx.lanl.gov/abs/0902.3519}{{\tt arXiv:0902.3519}}].

\bibitem{Heinrich:2009be}
G.~Heinrich, T.~Huber, D.~A. Kosower, and V.~A. Smirnov, {\it {Nine-Propagator
  Master Integrals for Massless Three-Loop Form Factors}},  {\em Phys.Lett.}
  {\bf B678} (2009) 359--366, [\href{http://xxx.lanl.gov/abs/0902.3512}{{\tt
  arXiv:0902.3512}}].

\bibitem{Gehrmann:2010ue}
T.~Gehrmann, E.~W.~N. Glover, T.~Huber, N.~Ikizlerli, and C.~Studerus, {\it
  {Calculation of the quark and gluon form factors to three loops in QCD}},
  {\em JHEP} {\bf 1006} (2010) 094,
  [\href{http://xxx.lanl.gov/abs/1004.3653}{{\tt arXiv:1004.3653}}].

\bibitem{Lee:2010ik}
R.~N. Lee and V.~A. Smirnov, {\it {Analytic Epsilon Expansions of Master
  Integrals Corresponding to Massless Three-Loop Form Factors and Three-Loop
  g-2 up to Four-Loop Transcendentality Weight}},  {\em JHEP} {\bf 1102} (2011)
  102, [\href{http://xxx.lanl.gov/abs/1010.1334}{{\tt arXiv:1010.1334}}].

\bibitem{Knizhnik:1984nr}
V.~Knizhnik and A.~Zamolodchikov, {\it {Current Algebra and Wess-Zumino Model
  in Two-Dimensions}},  {\em Nucl.Phys.} {\bf B247} (1984) 83--103.

\bibitem{Beneke:1997zp}
M.~Beneke and V.~A. Smirnov, {\it {Asymptotic expansion of Feynman integrals
  near threshold}},  {\em Nucl.Phys.} {\bf B522} (1998) 321--344,
  [\href{http://xxx.lanl.gov/abs/hep-ph/9711391}{{\tt hep-ph/9711391}}].

\bibitem{Smirnov:1998vk}
V.~A. Smirnov and E.~R. Rakhmetov, {\it {The Strategy of regions for asymptotic
  expansion of two loop vertex Feynman diagrams}},  {\em Theor.Math.Phys.} {\bf
  120} (1999) 870--875, [\href{http://xxx.lanl.gov/abs/hep-ph/9812529}{{\tt
  hep-ph/9812529}}].

\bibitem{Smirnov:1999bza}
V.~A. Smirnov, {\it {Problems of the strategy of regions}},  {\em Phys.Lett.}
  {\bf B465} (1999) 226--234,
  [\href{http://xxx.lanl.gov/abs/hep-ph/9907471}{{\tt hep-ph/9907471}}].

\bibitem{Smirnov:2002pj}
V.~A. Smirnov, {\it {Applied asymptotic expansions in momenta and masses}},
  {\em Springer Tracts Mod.Phys.} {\bf 177} (2002) 1--262.

\bibitem{Schutzmeier:2009}
T.~Schutzmeier, {\it {Matrix elements for the $\bar{B} \to X_s \gamma$ decay at
  NNLO. PhD Thesis, Universit\"at W\"urzburg, 2009}}, .

\bibitem{Bern:2007hh}
Z.~Bern, J.~J. Carrasco, L.~J. Dixon, H.~Johansson, D.~A. Kosower, and
  R.~Roiban, {\it {Three-Loop Superfiniteness of N=8 Supergravity}},  {\em
  Phys.Rev.Lett.} {\bf 98} (2007) 161303,
  [\href{http://xxx.lanl.gov/abs/hep-th/0702112}{{\tt hep-th/0702112}}].

\bibitem{Bogner:2013tia}
C.~Bogner and M.~Luders, {\it {Multiple polylogarithms and linearly reducible
  Feynman graphs}},  \href{http://xxx.lanl.gov/abs/1302.6215}{{\tt
  arXiv:1302.6215}}.

\bibitem{Brown:2008um}
F.~Brown, {\it {The Massless higher-loop two-point function}},  {\em
  Commun.Math.Phys.} {\bf 287} (2009) 925--958,
  [\href{http://xxx.lanl.gov/abs/0804.1660}{{\tt arXiv:0804.1660}}].

\bibitem{Panzer:2013cha}
E.~Panzer, {\it {On the analytic computation of massless propagators in
  dimensional regularization}},  \href{http://xxx.lanl.gov/abs/1305.2161}{{\tt
  arXiv:1305.2161}}.

\bibitem{Baikov:2010hf}
P.~Baikov and K.~Chetyrkin, {\it {Four Loop Massless Propagators: An Algebraic
  Evaluation of All Master Integrals}},  {\em Nucl.Phys.} {\bf B837} (2010)
  186--220, [\href{http://xxx.lanl.gov/abs/1004.1153}{{\tt arXiv:1004.1153}}].

\bibitem{Lee:2011jt}
R.~N. Lee, A.~V. Smirnov, and V.~A. Smirnov, {\it {Master Integrals for
  Four-Loop Massless Propagators up to Transcendentality Weight Twelve}},  {\em
  Nucl.Phys.} {\bf B856} (2012) 95--110,
  [\href{http://xxx.lanl.gov/abs/1108.0732}{{\tt arXiv:1108.0732}}].

\bibitem{Smirnov:2012gma}
V.~A. Smirnov, {\it {Analytic tools for Feynman integrals}},  {\em Springer
  Tracts Mod.Phys.} {\bf 250} (2012) 1--296.

\bibitem{Smirnov:2008iw}
A.~V. Smirnov, {\it {Algorithm FIRE -- Feynman Integral REduction}},  {\em
  JHEP} {\bf 0810} (2008) 107, [\href{http://xxx.lanl.gov/abs/0807.3243}{{\tt
  arXiv:0807.3243}}].

\bibitem{Smirnov:2013dia}
A.~V. Smirnov and V.~A. Smirnov, {\it {FIRE4, LiteRed and accompanying tools to
  solve integration by parts relations}},  {\em Comput.Phys.Commun.} {\bf 184}
  (2013) 2820--2827, [\href{http://xxx.lanl.gov/abs/1302.5885}{{\tt
  arXiv:1302.5885}}].

\bibitem{Remiddi:1999ew}
E.~Remiddi and J.~A.~M. Vermaseren, {\it {Harmonic polylogarithms}},  {\em Int.
  J. Mod. Phys.} {\bf A15} (2000) 725--754,
  [\href{http://xxx.lanl.gov/abs/hep-ph/9905237}{{\tt hep-ph/9905237}}].

\bibitem{Pak:2010pt}
A.~Pak and A.~V. Smirnov, {\it {Geometric approach to asymptotic expansion of
  Feynman integrals}},  {\em Eur.Phys.J.} {\bf C71} (2011) 1626,
  [\href{http://xxx.lanl.gov/abs/1011.4863}{{\tt arXiv:1011.4863}}].

\bibitem{Jantzen:2012mw}
B.~Jantzen, A.~V. Smirnov, and V.~A. Smirnov, {\it {Expansion by regions:
  revealing potential and Glauber regions automatically}},  {\em Eur.Phys.J.}
  {\bf C72} (2012) 2139, [\href{http://xxx.lanl.gov/abs/1206.0546}{{\tt
  arXiv:1206.0546}}].

\bibitem{Chetyrkin:1981qh}
K.~G. Chetyrkin and F.~V. Tkachov, {\it {Integration by Parts: The Algorithm to
  Calculate beta Functions in 4 Loops}},  {\em Nucl.Phys.} {\bf B192} (1981)
  159--204.

\bibitem{Lee:2013hzt}
R.~N. Lee and A.~A. Pomeransky, {\it {Critical points and number of master
  integrals}},  \href{http://xxx.lanl.gov/abs/1308.6676}{{\tt
  arXiv:1308.6676}}.

\bibitem{Smirnov:2010hn}
A.~V. Smirnov and A.~V. Petukhov, {\it {The Number of Master Integrals is
  Finite}},  {\em Lett.Math.Phys.} {\bf 97} (2011) 37--44,
  [\href{http://xxx.lanl.gov/abs/1004.4199}{{\tt arXiv:1004.4199}}].

\bibitem{Smirnov:1999gc}
V.~A. Smirnov, {\it {Analytical result for dimensionally regularized massless
  on shell double box}},  {\em Phys.Lett.} {\bf B460} (1999) 397--404,
  [\href{http://xxx.lanl.gov/abs/hep-ph/9905323}{{\tt hep-ph/9905323}}].

\bibitem{Tausk:1999vh}
J.~Tausk, {\it {Nonplanar massless two loop Feynman diagrams with four on-shell
  legs}},  {\em Phys.Lett.} {\bf B469} (1999) 225--234,
  [\href{http://xxx.lanl.gov/abs/hep-ph/9909506}{{\tt hep-ph/9909506}}].

\bibitem{FIESTA3}
A.~V. Smirnov, {\it {FIESTA 3: cluster-parallelizable multiloop numerical
  calculations in physical regions, 2013}}, .

\bibitem{Vermaseren:1998uu}
J.~Vermaseren, {\it {Harmonic sums, Mellin transforms and integrals}},  {\em
  Int.J.Mod.Phys.} {\bf A14} (1999) 2037--2076,
  [\href{http://xxx.lanl.gov/abs/hep-ph/9806280}{{\tt hep-ph/9806280}}].

\bibitem{Moch:2005uc}
S.~Moch and P.~Uwer, {\it {XSummer: Transcendental functions and symbolic
  summation in form}},  {\em Comput.Phys.Commun.} {\bf 174} (2006) 759--770,
  [\href{http://xxx.lanl.gov/abs/math-ph/0508008}{{\tt math-ph/0508008}}].

\bibitem{Czakon:2005rk}
M.~Czakon, {\it {Automatized analytic continuation of Mellin-Barnes
  integrals}},  {\em Comput. Phys. Commun.} {\bf 175} (2006) 559--571,
  [\href{http://xxx.lanl.gov/abs/hep-ph/0511200}{{\tt hep-ph/0511200}}].

\bibitem{Smirnov:2009up}
A.~V. Smirnov and V.~A. Smirnov, {\it {On the Resolution of Singularities of
  Multiple Mellin-Barnes Integrals}},  {\em Eur.Phys.J.} {\bf C62} (2009)
  445--449, [\href{http://xxx.lanl.gov/abs/0901.0386}{{\tt arXiv:0901.0386}}].

\bibitem{heptools1}
M.~Czakon, {\it {\tt MBasymptotics.m}},
  \href{http://xxx.lanl.gov/abs/http://projects.hepforge.org/mbtools/}{{\tt
  http://projects.hepforge.org/mbtools/}}.

\bibitem{heptools2}
D.~Kosower, {\it {\tt barnesroutines.m}},
  \href{http://xxx.lanl.gov/abs/http://projects.hepforge.org/mbtools/}{{\tt
  http://projects.hepforge.org/mbtools/}}.

\bibitem{Maitre:2005uu}
D.~Maitre, {\it {HPL, a mathematica implementation of the harmonic
  polylogarithms}},  {\em Comput.Phys.Commun.} {\bf 174} (2006) 222--240,
  [\href{http://xxx.lanl.gov/abs/hep-ph/0507152}{{\tt hep-ph/0507152}}].

\bibitem{Smirnov:2008py}
A.~V. Smirnov and M.~N. Tentyukov, {\it {Feynman Integral Evaluation by a
  Sector decomposiTion Approach (FIESTA)}},  {\em Comput. Phys. Commun.} {\bf
  180} (2009) 735--746, [\href{http://xxx.lanl.gov/abs/0807.4129}{{\tt
  arXiv:0807.4129}}].

\bibitem{Smirnov:2009pb}
A.~V. Smirnov, V.~A. Smirnov, and M.~Tentyukov, {\it {FIESTA 2: parallelizeable
  multiloop numerical calculations}},  {\em Comput. Phys. Commun.} {\bf 182}
  (2011) 790--803, [\href{http://xxx.lanl.gov/abs/0912.0158}{{\tt
  arXiv:0912.0158}}].

\bibitem{Brown:2012ia}
F.~Brown and O.~Schnetz, {\it {Proof of the zig-zag conjecture}},
  \href{http://xxx.lanl.gov/abs/1208.1890}{{\tt arXiv:1208.1890}}.

\bibitem{Drummond:2006rz}
J.~Drummond, J.~Henn, V.~Smirnov, and E.~Sokatchev, {\it {Magic identities for
  conformal four-point integrals}},  {\em JHEP} {\bf 0701} (2007) 064,
  [\href{http://xxx.lanl.gov/abs/hep-th/0607160}{{\tt hep-th/0607160}}].

\bibitem{Drummond:2010cz}
J.~M. Drummond, J.~M. Henn, and J.~Trnka, {\it {New differential equations for
  on-shell loop integrals}},  {\em JHEP} {\bf 1104} (2011) 083,
  [\href{http://xxx.lanl.gov/abs/1010.3679}{{\tt arXiv:1010.3679}}].

\bibitem{Broadhurst:1995km}
D.~J. Broadhurst and D.~Kreimer, {\it {Knots and numbers in $\phi^4$ theory to
  7 loops and beyond}},  {\em Int.J.Mod.Phys.} {\bf C6} (1995) 519--524,
  [\href{http://xxx.lanl.gov/abs/hep-ph/9504352}{{\tt hep-ph/9504352}}].

\end{thebibliography}\endgroup
 
\end{document}